\documentclass[apj]{emulateapj}

\usepackage{apjfonts}
\usepackage{natbib}
\usepackage{graphicx}
\usepackage{amssymb}
\usepackage{amsmath}
\bibpunct{(}{)}{;}{a}{}{,}

\newcommand{\mni}[1]{{Mn\,{\sc i}\,$\lambda$}#1}
\newcommand{\mnis}[1]{{Mn\,{\sc i}}#1}
\newcommand{\feis}[1]{{Fe\,{\sc i}}#1}
\newcommand{\mniis}[1]{{Mn\,{\sc ii}}#1}

\epsscale{0.5}

\begin{document}
\title{Quiet Sun Magnetic Field Measurements Based on Lines with Hyperfine 
Structure}

\author{J.~S\'anchez~Almeida\altaffilmark{1},
	B. Viticchi\'{e}\altaffilmark{2},
	E.~Landi~Degl'Innocenti\altaffilmark{3},
	and
	F.~Berrilli\altaffilmark{2}}
\altaffiltext{1}{Instituto de Astrof\'\i sica de Canarias, E-38205 La Laguna, Tenerife, Spain}
\altaffiltext{2}{Dipartimento di Fisica, Universit\`a di Roma {\em Tor Vergata}, 
	Via della Ricerca Scientifica 1, I-00133 Rome, Italy} 
\altaffiltext{3}{Universit\`a degli Studi di Firenze, Dipartimento di Astronomia e Scienza dello Spazio, 
	Largo E. Fermi 2, 50125 Firenze, Italy}

\email{jos@iac.es,bartolomeo.viticchie@roma2.infn.it,\\
	landie@arcetri.astro.it,
	francesco.berrilli@roma2.infn.it}
%

\begin{abstract}
The Zeeman pattern of Mn~{\sc i} lines is 
sensitive to  hyperfine structure (HFS) and, because of this
reason, they 
respond to hG magnetic field strengths 
differently from the lines commonly used in solar 
magnetometry. This peculiarity has been employed
to measure magnetic field strengths in quiet Sun 
regions. However, the methods applied so far assume the 
magnetic field to be constant in the resolution element. 
The assumption
is clearly insufficient to describe the 
complex quiet Sun magnetic fields, biasing 
the results of the measurements. The diagnostic
potential of Mn~{\sc i} lines can be fully exploited
only after understanding the sense and the magnitude
of such bias. We present the first syntheses of Mn~{\sc i} 
lines in realistic quiet Sun model atmospheres. 
Plasmas varying in magnetic field strength, magnetic field 
direction, and  
velocity, contribute to the synthetic polarization signals.
The syntheses show how the Mn~{\sc i} lines weaken with increasing 
field strength. In particular, kG magnetic concentrations
produce \mni{5538} circular polarization signals 
(Stokes~$V$) which can be up to two orders of magnitude
smaller than what the weak magnetic field approximation 
predicts. As corollaries of this result, (1) the polarization emerging 
from an atmosphere having weak and strong fields is 
biased towards the weak fields, and (2) HFS features 
characteristic of weak fields show up even when the magnetic 
flux and energy are dominated by kG fields. For the HFS feature of \mni{5538} to 
disappear the filling factor of kG fields has to be larger than 
the filling factor of sub-kG fields. 
Since the Mn~{\sc i} lines are usually weak, Stokes~$V$ depends on 
magnetic field inclination 
according to the simple consine law, a scaling that
holds independently of the 
magnetic field strength. 
Atmospheres with unresolved velocities produce 
very asymmetric line profiles, which cannot be 
reproduced by simple one-component model atmospheres.
Inversion techniques incorporating complex magnetic 
atmospheres must be implemented for a proper diagnosis. 
Using the HFS constants available in the literature we 
reproduce the observed line profiles of nine
lines with varied HFS patterns.
Consequently, the uncertainty of the HFS constants
do not seem to limit the use of Mn~{\sc i} lines for 
magnetometry.
\end{abstract}


\maketitle
\section{Introduction}
\label{intro}

When the polarimetric sensitivity and the angular resolution
exceed the required threshold, magnetic signals show up 
almost everywhere on the solar photosphere. 
The signals in supergranulation cell interiors
are particularly weak, however, most
of the solar surface is in the form of cell interiors
and, therefore, these weak signals probably set the total
unsigned magnetic flux and magnetic energy existing
in the photosphere at any given time 
\citep[e.g.,][]{unn59,ste82,yi93,san98c,san04,sch03b}. 
The importance of these ubiquitous
quiet Sun magnetic fields 
depends, to a large extent, on the magnetic field 
strengths characterizing them. For example, the magnetic flux and
the magnetic energy scale as powers of the field strength, 
and the connectivity between photosphere and
corona is probably provided by the  
magnetic concentrations with the largest field strengths
\citep[e.g.,][]{dom06}. Unfortunately, measuring
quiet Sun magnetic field strengths is not a trivial
task.
All measurements are based on the residual polarization 
left when a magnetic field of complex topology is 
observed with finite angular resolution 
\citep[e.g.,][]{emo01,san03,tru04}. Measuring implies assuming 
many properties of the unresolved complex magnetic 
field and, in doing so, the measurements become model dependent
and prone to bias. Depending on the
technique used for measuring, the real quiet Sun exhibits 
weak daG fields \citep[e.g.,][]{ste82,fau93,bom05}, intermediate hG fields 
\citep[e.g.,][]{lin99,kho02,lop06}, or strong kG fields 
\citep[e.g.,][]{san00,lit02,dom03b}. 
Such discrepancies can be naturally understood if 
the true quiet Sun contains a continuous distribution
of field strengths going all the way from zero to two kG. 
Different techniques are biased differently and,
therefore, they tend to pick out a particular part of the 
distribution. 
This scenario is very much consistent with realistic
numerical simulations of magneto-convection
\citep[][]{cat99a,stei06,vog05,vog07}.
In order to provide a comprehensive observational
description of the quiet Sun magnetic field
strengths, one has to combine different 
methods carefully chosen to have complementary biases. Then 
the full distribution can be assambled taking 
the biases into account \cite[][]{dom06}.

In an effort to complement the existing magnetic field
strength diagnostic techniques, \citet{lop02} proposed the 
use of spectral lines whose Zeeman patterns are sensitive 
to hyperfine structure (HFS).  
The formalism to deal with the HFS of spectral lines in magnetic 
atmospheres was developed more than thirty years ago by \citet{lan75}. 
According to such formalism, the polarization of the HFS lines vary 
with magnetic field strength  very differently from the lines 
commonly used in solar magnetometry. This unusual behavior was
invoked by \citet{lop02} when proposing the use of
HFS Mn~{\sc i}  lines as a  diagnostic tool for magnetic 
field strengths. L\'opez Ariste and coworkers 
have applied the idea to measure magnetic field strengths in 
quiet Sun regions \citep{lop06,lop07,ase07}.
Since the number of observables is limited,
they minimize the statistical error of the measuremet
by  minimizing the number of free parameters 
to be tuned. Milne-Eddington atmospheres (ME)
are used to synthesize the polarization of the lines. 
The magnetic field is assumed to be 
constant and, therefore, the measurements provide some kind of 
weighted average of the true field strengths existing in the 
resolution  elements. As we pointed out above, the
topology of the quiet sun magnetic field is complex, 
with a distribution of field strengths and polarities coexisting 
in a typical resolution element. Then the ill-defined
average provided by the Mn~{\sc i} lines is expected to be 
biased toward a particular range of field strengths, as it happens 
with the rest of the quiet Sun measurements 
(e.g., the NIR Fe~{\sc i} lines overlook kG magnetic 
	concentrations, \citealt{san00}, \citealt{soc03}; 
	the traditional visible Fe~{\sc i} lines exaggerate
	the contribution of kG fields, \citealt{bel03}, \citealt{mar06};
	the Hanle scattering polarization signals are not sensitive 
	to hG and kG, \citealt{fau01}, \citealt{san05}).
The bias presented by the HFS Mn~{\sc i} lines is so far unknown, 
and the existing and forthcoming measurements based on those lines
will be fully appreciated only when the sources 
of systematic effects are properly acknowledged and quantified.
In order to explore the magnitude and the sense of the 
expected effects, 
we undertake the synthesis of  Mn~{\sc i} lines in a number of realistic 
quiet Sun atmospheres with complex magnetic field distributions. 
The main trends are presented here.

The paper is organized as follows. 
Section~\ref{code} describes the software developed
to carry out the syntheses. First, a ME code
is needed to compare our syntheses with the results
in the literature. Then, a plane parallel one-dimensional (1D) code allows
us to explore the influence of realistic thermodynamic
conditions on the polarization of lines with HFS.
Finally, a MIcro-Structured Magnetic Atmosphere
(MISMA) code provides additional realism to the
modeling since it includes  coupling 
between magnetic field strengths and thermodynamic
conditions, magnetic fields varying along and
across the line-of-sight (LOS),  
mixed polarities in the resolution element,
etc. 
All these codes are based on the original FORTRAN
routine written by \citet{lan78}.
The analysis is focused on the line most often
used in 
observations, namely, \mni{5538}.
We discuss its intensity and circular
polarization, since the linear polarization signals
are very weak and they  
remain  undetected so far.
Single component and multi-component syntheses 
of this line are described in 
\S~\ref{scs} and \S~\ref{mcs}, respectively.
An exploratory attempt to consider unresolved mixed polarities 
and velocity fields is carried out in \S~\ref{comp_obs}.
The polarization of other Mn~{\sc i} lines
is also considered in \S~\ref{other}. 
The implications of these syntheses are 
discussed in \S~\ref{conclusions}.

\section{Description of the synthesis procedures}\label{code}
The synthesis of Stokes profiles has been carried out
using three different approaches to solve the radiative 
transfer equation for polarized light,
\begin{equation}
\label{rte}
\frac{d\textbf{I}}{ds}=-\textbf{K}(\textbf{I}-\textbf{S}),
\end{equation}
where $\textbf{I}=(I,Q,U,V)^\dag$ is the column vector containing the Stokes parameters, 
$\textbf{S}=(B_T,0,0,0)^\dag$ is the source vector with $B_T$ the LTE source function, 
$\textbf{K}$ is the $4\times4$ absorption matrix and, finally,  the variable of
integration $s$ corresponds to the length along the LOS. 
(For details on matrix elements and the rest of the equation, see
\citeauthor{lan76}~\citeyear{lan76}.)
The three codes differ in the model atmosphere used to describe the photosphere. 
The first one solves equation~(\ref{rte}) under ME hypothesis. 
The source function is assumed to vary linearly  with the continuum 
optical depth $\tau$. Then
the radiative transfer equation admits an analytic solution involving the
two coefficients of the source function ($B_T=B_0+B_1\tau$) and the seven 
coefficients of the absorption matrix, which are assumed to be
constant (see \citeauthor{lan75}~\citeyear{lan75}; equation [14]).
The second code solves equation~(\ref{rte}) for one-dimensional (1D) atmospheres.
Realistic temperatures, pressures and magnetic field vectors are considered,
but they only vary along the LOS so that a single ray suffices 
to synthesize the spectrum. 
The third code solves a different version of equation~(\ref{rte}). It is obtained by
a spatial averaging under the MISMA hypothesis 
\citep[][]{san96}. 
It assumes the solar atmosphere to have inhomogeneities 
smaller than the photon mean-free-path
whose details are washed out by the radiative transfer process. 
Only their average properties matter.
The new equation to be solved is,
\begin{equation}
\label{rtem}
\frac{d\textbf{I}}{ds} = -\left\langle \textbf{K} \right\rangle(\textbf{I}-\textbf{S}'),
\end{equation}
where $\textbf{S}'= \left\langle \textbf{K} \right\rangle^{-1} \left\langle \textbf{KS} \right\rangle$ is a 
mean source vector.
The brackets in equation~(\ref{rtem}) denote average over a volume whose  
dimension along the LOS is of the order of the photon mean-free-path. 
As a result of the averaging, $\left\langle \textbf{K} \right\rangle $ and $\textbf{S}'$ are 
slowly varying functions of $s$, with this slow variation 
accounting for the large scale structure in the atmosphere.
Pursuing the idea that optically-thin details are irrelevant for the final 
spectrum, \citet{san97b} puts forward a type of model MISMA made of distinct 
components with different temperatures, 
pressures and  magnetic field vectors.  The coexistence of different
components imposes a number of physical constraints to be 
satisfied, in particular, the lateral pressure balance 
causes the magnetic field to be coupled with the thermodynamics
(see the original reference for an exhaustive description).
These model MISMAs, with their constraints, provide 
the degree of realism required to carry out our exploratory 
work. Note that the model MISMAs reproduce all kinds of Stokes profiles observed 
in the quiet Sun, including those with extreme asymmetries 
\citep{san00,soc02,dom06b}. Equation~(\ref{rte}) is formally identical to 
equation~(\ref{rtem}). 
They are integrated using a predictor-corrector method with the variables equi-spaced in 
a fixed grid of atmospheric heights.
As we mention in the introduction, the linear polarization signals
$Q$ and $U$ are very weak, and we do not analyze them here despite 
the fact that they are synthesized together with $I$ and~$V$.

All the codes are based on the FORTRAN procedure HYPER by \citet{lan78} that 
computes the Zeeman pattern of any line with hyperfine structure.
The pattern depends on the
externally imposed magnetic field, together with a set
of atomic parameters, namely, the hyperfine structure constants, 
the quantum numbers of the two levels involved in the transition, 
the relative isotopic abundance, and the isotope shifts.
When the Zeeman splitting is comparable to the hyperfine structure 
splitting, the Stokes profiles are strongly disrupted 
showing typical hyperfine structure features. The 
pattern 
depends on the so-called {hyperfine structure constants}
$\mathcal{A}$ and $\mathcal{B}$, which account for the 
two first terms of the Hamiltonian describing the 
interaction between the electrons in an atomic level and the 
nuclear magnetic moment --  $\mathcal{A}$ corresponds to
the magnetic-dipole coupling whereas 
$\mathcal{B}$ describes the electric-quadrupole coupling. 
The constants  $\mathcal{A}$ and $\mathcal{B}$ can be determined empirically or 
theoretically. 
The values of $\mathcal{A}$ adopted for our Mn~{\sc i} syntheses 
are listed in Table \ref{param}. The table includes the appropriate
references, which indicate that $\mathcal{B}$ is negligible small in all cases. 
\begin{deluxetable*}{ccccccccc}
\tablecolumns{9}
\tablewidth{0pc}
\tabletypesize{\footnotesize}
\tablecaption{Atomic parameters used in the synthesis of \mnis{} lines\label{param}}
\tablehead{
\colhead{Line} & \colhead{Wavelength\tablenotemark{a}} & \colhead{Transition\tablenotemark{a}}
 & \colhead{$\chi$\tablenotemark{a}} & \colhead{$\log gf$} & \colhead{$g_{\rm low}$\tablenotemark{a}} 
& \colhead{$g_{\rm up}$\tablenotemark{a}} & \colhead{$\mathcal{A}_{\rm low}$} & \colhead{$\mathcal{A}_{\rm up}$}\\
\colhead{} & \colhead{[\AA]} & \colhead{} & \colhead{[eV]} & \colhead{} & \colhead{} & \colhead{} & \colhead{[$10^{-3}$cm$^{-1}$]} & \colhead{[$10^{-3}$cm$^{-1}$]}
}
\startdata
\mni{4457} & 4457.04 & $z^6P^0_{5/2}-e^6D_{3/2}$ & $3.073$ & $-0.555$\tablenotemark{a} & $1.875$ & $1.759$ & $15.6$\tablenotemark{b} & $22.8$\tablenotemark{b} \\ 
\mni{4470} & 4470.14 & $a^4D_{3/2}-z^4D^0_{3/2}$ & $2.940$ & $-0.444$\tablenotemark{a} & $1.198$ & $1.200$ & $1.7$\tablenotemark{c} & $6.4$\tablenotemark{c} \\ 
\mni{4502} & 4502.22 & $a^4D_{5/2}-z^4D^0_{7/2}$ & $2.919$ & $-0.344$\tablenotemark{a} & $1.368$ & $1.427$ & $-4.6$\tablenotemark{c} & $1.5$\tablenotemark{c} \\ 
\mni{4739} & 4739.11 & $a^4D_{3/2}-z^4F^0_{3/2}$ & $2.940$ & $-0.490$\tablenotemark{a} & $1.198$ & $0.400$ & $1.7$\tablenotemark{c} & $22.3$\tablenotemark{c} \\ 
\mni{5420} & 5420.36 & $a^6D_{7/2}-y^6P^0_{5/2}$ & $2.142$ & $-1.460$\tablenotemark{a} & $1.584$ & $1.886$ & $13.5$\tablenotemark{d} & $-21.5$\tablenotemark{d} \\ 
\mni{5516} & 5516.77 & $a^6D_{3/2}-y^6P^0_{3/2}$ & $2.178$ & $-1.847$\tablenotemark{a} & $1.866$ & $2.400$ & $16.5$\tablenotemark{d} & $-31.5$\tablenotemark{d} \\ 
\mni{5538} & 5537.76 & $a^6D_{1/2}-y^6P^0_{3/2}$ & $2.186$ & $-1.920$\tablenotemark{e} & $3.327$ & $2.400$ & $27.8$\tablenotemark{d} & $-31.5$\tablenotemark{d} \\
\mni{8741} & 8740.93 & $y^6P^0_{7/2}-e^6D_{9/2}$ & $4.434$ & $-0.05$~\tablenotemark{f} & $1.712$ & $1.554$ & $-13$\tablenotemark{b} & $15.5$\tablenotemark{b}\\
\mni{15262} & 15262.70\tablenotemark{g} & $e^8S_{7/2}-y^8P^0_{5/2}$~\tablenotemark{g} & $4.889$ & $+0.45$~\tablenotemark{f} & $2.000$ & $2.284$\tablenotemark{g} 
	& $25.2$\tablenotemark{b} & $27.5$\tablenotemark{b}
\enddata
\tablenotetext{a}{NIST Atomic Spectra Database, \citet{nist07}}
\tablenotetext{b}{\citet{Lefe03}}
\tablenotetext{c}{\citet{Blac05}}
\tablenotetext{d}{\citet{FisP39}}
\tablenotetext{e}{\citet{Marg72}}
\tablenotetext{f}{Set to match the observed solar spectrum with $\log$[Mn]=5.31.}
\tablenotetext{g}{\citet{ase07}}
\tablecomments{
The symbols undefined in the main text
stand for:  lower-level  oscillator strength $gf$,
  lower-level Lande factor $g_{\rm low}$, 
 upper-level Lande factor $g_{\rm up}$, and
 lower-level excitation potential $\chi$.
}
\end{deluxetable*}

\subsection{Testing the codes}
\label{test}
We carry out different types of tests using the line \mni{5538} as target.
Intensity profiles are compared with the unpolarized 
solar atlas. The syntheses are also compared with the (ME) Stokes profiles 
existing in the  literature. The codes are cross-checked one with respect to another, 
and with existing synthesis codes that neglect the HFS.
All these efforts are summarized next.

\begin{figure}
\plotone{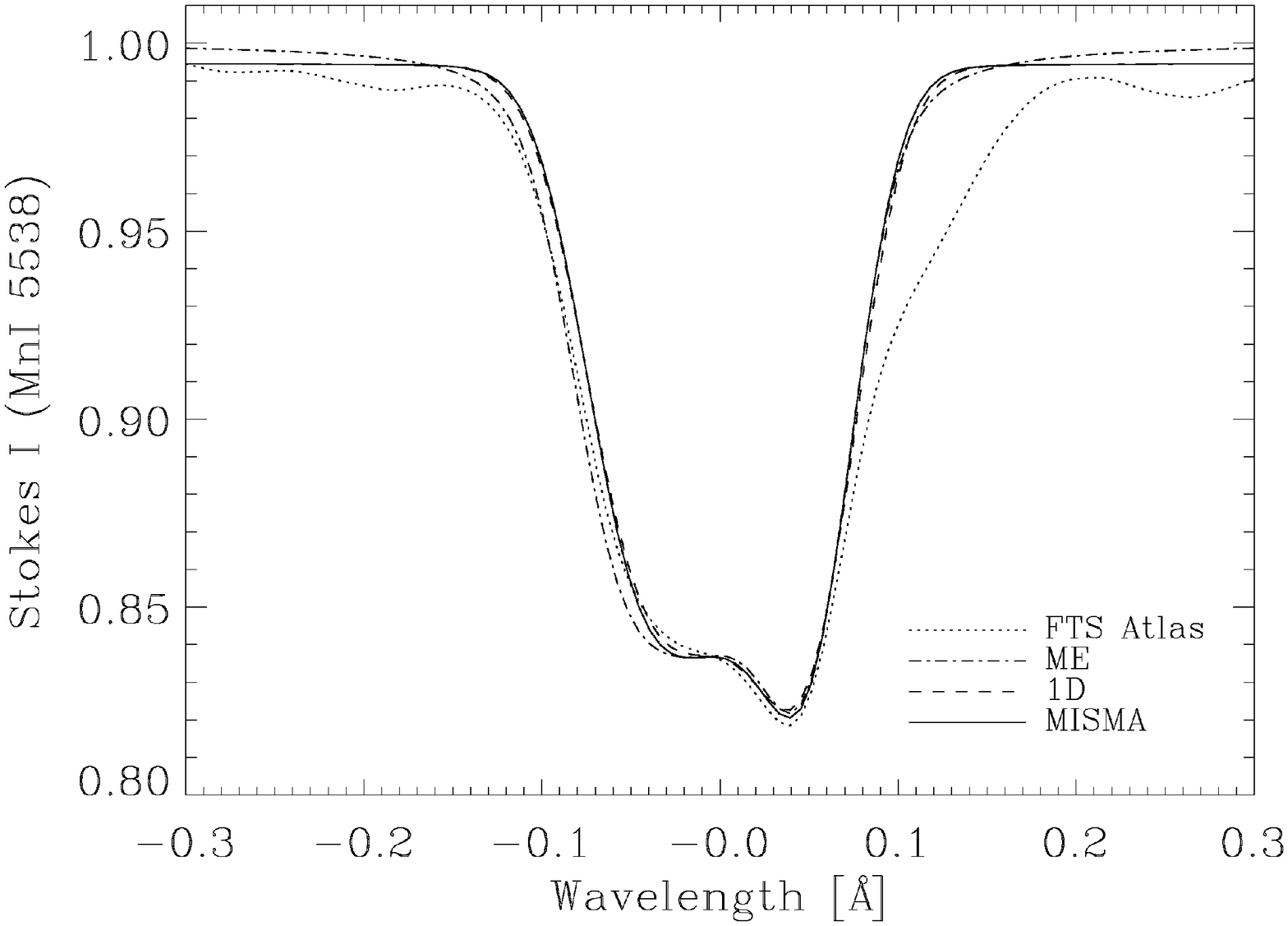}
\caption{\mni{5538} Stokes~$I$ profiles synthesized using the three 
codes and compared with the profile from the solar FTS atlas. The inset identifies
the different codes and the observation.
The wavelengths, in units of \AA ,  are referred to the laboratory 
wavelength of the line.
The profiles are normalized to the quiet Sun continuum intensity.
}
\label{si}
\end{figure}

The first series of tests is based on non-magnetic atmospheres.
We 
start by recovering 
the ME Stokes~$I$ profile in \citet[][Fig.3]{lop02} when using 
the appropriate ME parameters (L\'opez Ariste~2007, private communication). 
In addition,
Fig.~\ref{si} shows a reasonable ME fit to the
FTS solar spectrum \citep{nec99} using a Doppler width
$\Delta\lambda_D=0.029$~m\AA , a damping constant $a=0.3$, an absorption
coefficient $\eta_0=2.7$, a source function given by
$B_0=1.0$ and $B_1=1.1$,
and the atomic parameters in Table~\ref{param}.
Figure~\ref{si} also includes a fit to the observed profile 
using the 1D code. The line is synthesized in the
quiet Sun model atmosphere by \citet{mal86} using 
the atomic parameters in Table~\ref{param}. The fit was carried out
by trial-and-error varying the velocity along the LOS,
the macroturbulence ($v_{mac}$, set to 1.3~km~s$^{-1}$),  
and the Mn abundance ($\log[$Mn$]$, set to 5.30 in the  scale where $\log[$H$]$=12).
The MISMA code was also tested against the observed Stokes~$I$ assuming the thermodynamic
parameters of the model atmosphere to be given 
by the model quiet Sun in 
\citet[][\S~4.1]{san97b}.
As in the case of 1D atmospheres, the atomic parameters are those listed
in Table~\ref{param}, and we vary the velocity along the LOS,
the macroturbulence ($v_{mac}=$1.5~km~s$^{-1}$), 
and the Mn abundance ($\log[$Mn$]$=5.31). 
One additional consistency test was carried out with the 1D code. We compare the synthesis
obtained neglecting the hyperfine structure (i.e., imposing $\mathcal{A}=\mathcal{B}=0$) 
with the intensity of the 1D code from which we inherited the 
absorption coefficient 
routines \citep[][\S~5]{san92a}. The profiles of the two syntheses
are identical within the numerical precision.

\begin{figure}
\plotone{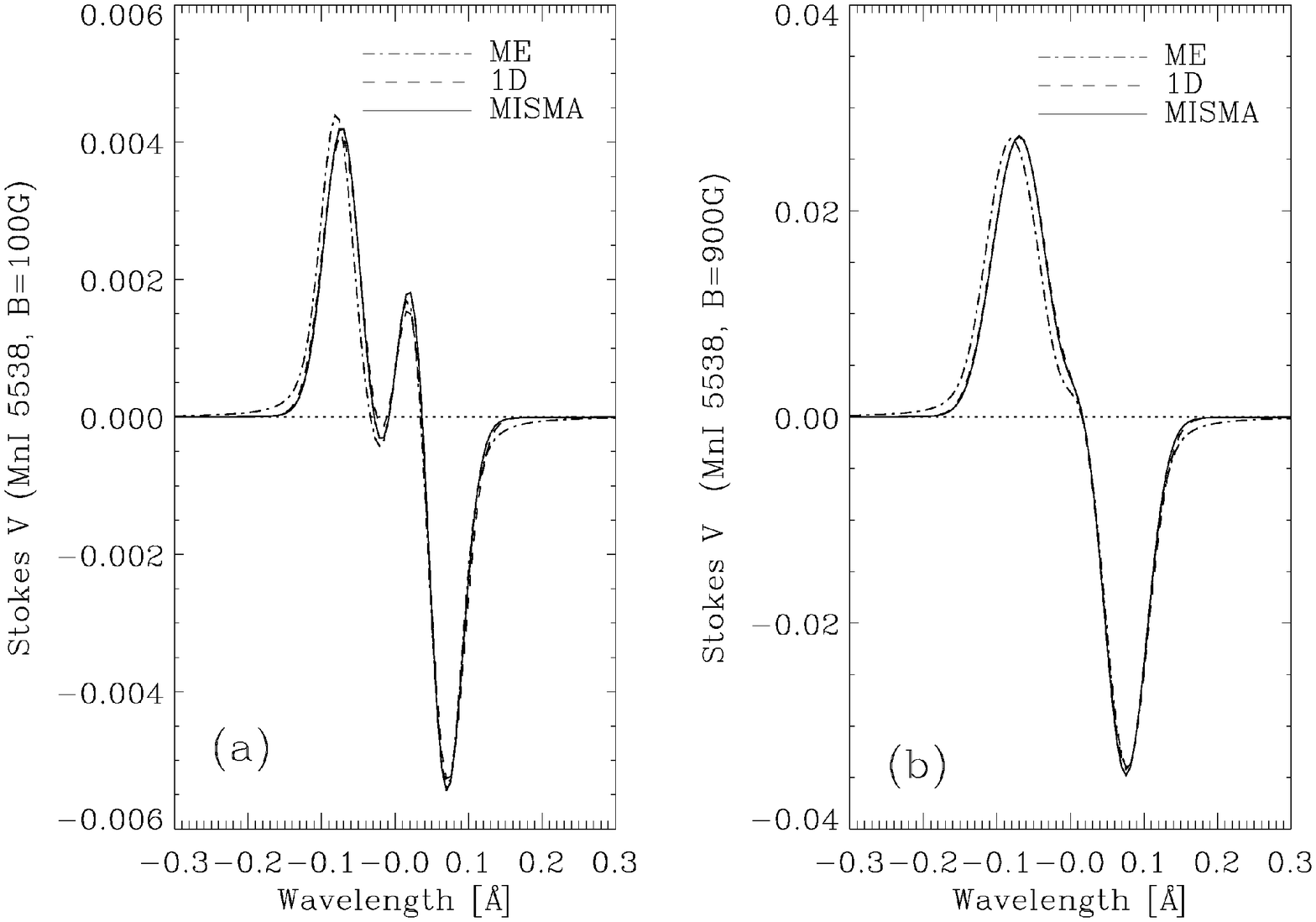}
\caption{\mni{5538} Stokes~$V$ profiles synthesized using the three
codes with $B=100~$G (a) and $B=900~$G (b). The insets 
pair each type of line with the corresponding
synthesis code.
The profiles are normalized to the quiet Sun continuum intensity,
and the wavelengths are refereed to the laboratory wavelength
of the line.
}
\label{svb}
\end{figure}

The second series of tests refers to magnetic atmospheres.
Employing the atomic parameters and the atmospheres described above, 
we include  a constant magnetic field vector. We use three different magnetic field strengths,
$B=100~$G, 300~G, and 900~G, with the same inclination, 
$\theta=60^\circ$, and azimuth, $\phi=0^\circ$.
The ME syntheses reproduce the Stokes profiles in Fig.~4 of \citet[][]{lop02}.
Stokes~$V$ profiles synthesized with the three codes 
 are shown in Fig.~\ref{svb}. 
The agreement between the syntheses supports their internal 
consistency  -- if the codes are tuned to
produce similar Stokes~$I$ profiles (Fig.~\ref{si}), 
then 
for the same field strength
they also produce similar Stokes~$V$ (Fig.~\ref{svb}).
The Stokes~$V$ profiles represented in Fig.~\ref{svb} correspond to (and are consistent
with) the two behaviors of \mni{5538} described by \citet[][]{lop02}. 
The left panel in Fig.~\ref{svb} 
shows the kind of profile for $B \lesssim 700~$G, with a characteristic reversal not far from the 
line core (hereafter HFS hump). Figure~\ref{svb}b corresponds 
to $B \gtrsim 700~$G, which do not generate HFS humps at the core. The presence 
or absence of such reversal is used by L\'opez~Ariste and co-workers
to distinguish between sub-kG fields 
(present) and kG fields (absent).

The results in \S~\ref{other} can be regarded as a third independent test. 
The  same non-magnetic quiet Sun model MISMA  that
reproduces the observed intensity of \mni{5538} 
also accounts
for eight additional \mnis{} lines with very different HFS patterns. 
Such agreement shows the realism and 
self-consistency of the syntheses.

\section{Syntheses of \mni{5538} in 1D atmospheres}\label{scs}

\begin{figure}
\plotone{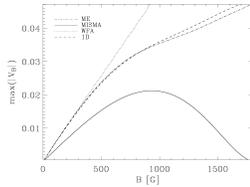}
\caption{Maximum Stokes~$V$ amplitude as a function of the 
magnetic field strength for ME, 1D, and MISMA model atmospheres. 
The prediction corresponding to the weak field approximation 
is represented as the dotted line. All magnetic field 
inclinations are set to $60^\circ$. Magnetic field strengths
are given in G.
}
\label{maxv}
\end{figure}

The weak magnetic field approximation 
is routinely used in solar 
magnetometry to, e.g., calibrate magnetograms.
It is also valid for lines with HFS
\citep{lan75,lan04}, and so it was
used by \citet{lop06} to estimate the
relative contribution of weak and strong fields to the observed
quiet Sun signals. 
When the approximation holds the Stokes~$V$ profile is  
proportional to the longitudinal component of the magnetic field
$B\cos\theta$,
\begin{equation}
\label{wfa}
V = - c\ {{dI}\over{d\lambda}} B\, \cos\theta,
\end{equation}
where the constant $c$ depends on the particular
spectral line, and the derivative
of the intensity profile, ${{dI}/{d\lambda}}$, gives
the variation with wavelength of Stokes~$V$\footnote{
Note, incidentally, that given the observed intensity profile of 
\mni{5538} (Fig.~\ref{si}), a hump at its 
Stokes~$V$ line core
is unavoidable when $B\rightarrow 0$.
}
\citep{lan75,lan04}.
We tested the approximation with the quiet Sun model atmospheres described in 
\S~\ref{test}, which 
are realistic in the sense that they
reproduce the 
observed intensity profile when the magnetic field strength
is close to zero (Fig.~\ref{si}).
The syntheses assume the magnetic field to be constant along the 
LOS, with the inclination $\theta$ set to 60$^{\circ}$. 
Figure~\ref{maxv} shows the variation of the
maximum Stokes~$V$ signal as a function of the magnetic field
strength. 
The weak field approximation breaks down for fairly low
magnetic field strengths, i.e., 
$B \gtrsim 400~$G (compare the dashed line
and the dotted line in Fig.~\ref{maxv}). Even more, 
the deviations are very important 
for kG magnetic field strengths. 
When the field strength 
is 1.5~kG the weak field approximation yields a Stokes~$V$ signal
twice larger than the synthetic signals in 1D model atmospheres
(Fig.~\ref{maxv}). 
In order to test the dependence  on $\cos\theta$ predicted
by equation~(\ref{wfa}),
we also carried out syntheses with various magnetic field inclinations. 
This dependence 
turns out to be closely followed by the synthetic signals, even for strong kG
magnetic field strengths. The behavior was to be expected since
\mni{5538} is a weak line satisfying the weakly polarizing medium 
approximation \citep{san99b}. In this approximation Stokes~$V$ 
scales with its specific absorption 
coefficient, which is proportional to $\cos\theta$ independently
of whether or no the spectral line has HFS.

\begin{figure}
\plotone{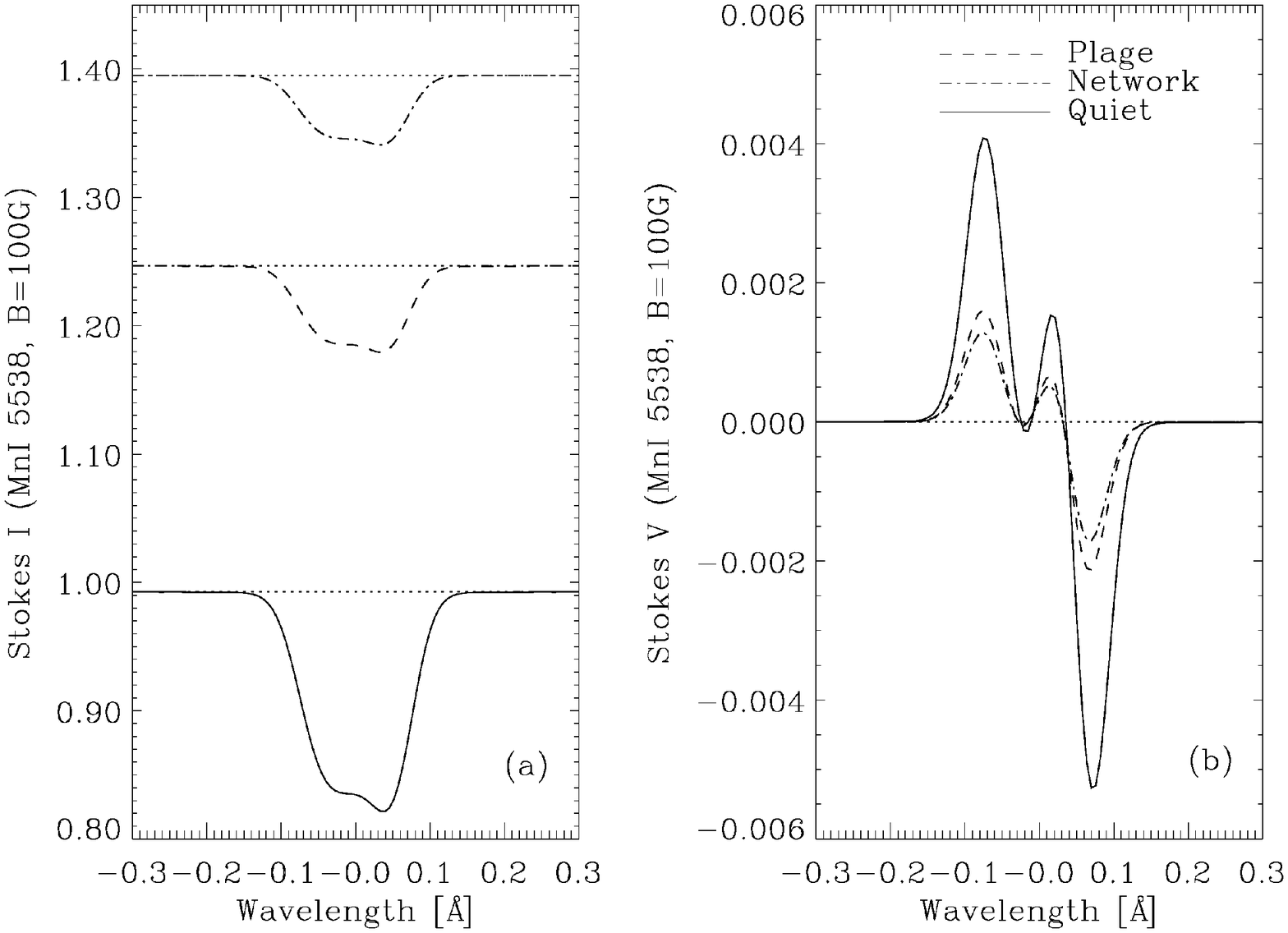}
\caption{Stokes~$I$ and $V$ profiles synthesized in three 1D model atmospheres
corresponding to the quiet Sun (the solid line), a network region
(the dotted dashed line), and a plage region (the dashed line). 
The magnetic field is constant with a strength of 100~G and an
inclination of 60$^\circ$. All profiles have been normalized to the quiet Sun
continuum intensity. The dotted line is just a reference to show the
continuum level.}
\label{slk}
\end{figure}

\begin{figure}
\plotone{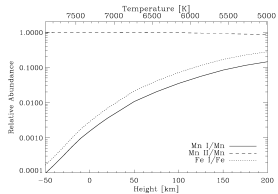}
\caption{Relative abundance of \mnis{} and \mniis{} as a function of the
height in the atmosphere for a quiet Sun stratification of temperatures. 
(The temperature corresponding to each height appears on the top of the figure.) 
The gas pressure is the same for all the heights, and it has been chosen to match the 
gas pressure of the non-magnetic quiet Sun some $100~$km above the continuum optical 
depth equals one.
The \feis{} abundance is included for comparison -- 
its overabundance with respect to \mnis{} 
is due to its higher ionization potential.
}
\label{mnimnii}
\end{figure}

The quiet Sun contains plasmas with all magnetic field
strengths from 0 to 2~kG (\S~\ref{intro}). According to a mechanism 
originally put forward by \citet{spr76} and now well established, 
one expects a strong correlation between the magnetic field strength and 
the temperature of the observable photospheric 
layers. Plasmas having kG fields must be 
strongly evacuated to stay in mechanical balance within the
photosphere. Consequently, strongly magnetized plasmas are transparent,
showing light coming from
deep and therefore hot (sub-)photospheric layers. 
In order to explore the dependence of the \mni{5538} Stokes~$V$ signals
on the thermodynamics, we compare synthetic
profiles formed under different thermodynamic conditions
but the same magnetic field strength. 
The line is synthesized in a quiet Sun model atmosphere \citep{mal86}, 
a network model atmosphere \citep{sol86,sol88b}, and a plage model atmosphere 
\citep{sol86,sol88b}, with the temperature increasing from quiet Sun to network.
The results are shown in Fig.~\ref{slk}.
First note that the continuum intensity of Stokes~$I$ increases from quiet Sun (the
coolest model) to network (the hotest model). In anticorrelation with
continuum intensity, the Stokes~$V$ signals decrease with increasing temperature
(Fig.~\ref{slk}, right). 
This behavior is not attributable to the HFS,
but it is due to the ionization balance of Mn in the
photosphere. Mn~{\sc i} is the minor species
so that a small modification of the ionization
equilibrium does not alter the Mn~{\sc ii} abundance but it drastically changes
the number of Mn~{\sc i} atoms available for absorption. 
Increasing the temperature increases the ionization, and the  Mn~{\sc i} lines
weaken. 
Figure~\ref{mnimnii} shows the relative abundance of \mnis\ as a function of height 
in the atmosphere for a constant density representative of the  
layers where the lines are formed (say, 100~km above continuum optical
depth equals one). According to Fig.~\ref{mnimnii}, if the presence of a strong 
magnetic field decreases the formation height of a spectral line by $100~$km, 
then the \mnis{} is depleted by more than one order of magnitude 
(compare the values at 100~km and 0~km). 
Note that this behavior is common to
all \mnis{} lines, and it is also typical of other minor species
like  Fe~{\sc i} 
(e.g., the dotted line in Fig.~\ref{mnimnii}).
In short, the \mnis{} lines are expected to weaken with increasing magnetic field 
strength.

The coupling between magnetic field strength and temperature 
is fully accounted for in the MISMA syntheses, which partly 
explains the dimming of the kG Stokes~$V$ signals shown in Fig.~\ref{maxv} 
(the solid line). Another part of the dimming is due to non-magnetic plasma 
obscuring the magnetic plasma, and effect specific to MISMAs
and elaborated by \citet{san00c}. 

%
%
\section{Multicomponent synthesis of \mni{5538}}
\label{mcs}

The syntheses carried out so far assume very simple magnetic 
configurations. The magnetic field is either constant or, in the case 
of model MISMAs, one magnetic component varies along the LOS. 
Due to the complex topology of the quiet Sun magnetic fields,
our limited angular resolution, and the spatial averages carried out to
obtain 
appropriate 
signals, those simplifications are insufficient
to represent observed Stokes profiles. Real polarization signals are
formed in plasmas having a range of magnetic properties, therefore, 
we improve the realism of the synthesis assuming multi-component 
model atmospheres. Specifically, we assume
\begin{equation}
I=\int_{0}^{B_{\rm max}}{P(B)\,I_B\,dB}, \label{pdfIV}
\end{equation}
and
\begin{equation}
V=\int_{0}^{B_{\rm max}}{\langle\cos\theta\rangle\,P(B)\,V_B\,dB}. \label{pdfIVb}
\end{equation}
The symbols $I_B$ and $V_B$ represent the Stokes $I$ and $V$ profiles 
produced by an atmosphere with a single longitudinal magnetic field 
of strength $B$, whereas $P(B)$ stands for 
the fraction of resolution element occupied by such atmosphere.
$P(B)$ is usually referred to as magnetic field
strength probability density function (PDF). It 
is normalized to one, and it can be envisaged as the
filling factor corresponding to each $B$. 
The maximum field strength existing in the quiet Sun 
sets $B_{\rm max}$ to some 2~kG.
As we show in Appendix~\ref{appa}, equations~(\ref{pdfIV}) 
and (\ref{pdfIVb}) hold even for an arbitrary distribution
of magnetic field directions, 
provided that the average inclination factor 
$\langle\cos\theta\rangle$ is computed properly. 
We would like to make it clear, however, that
the PDF approach for representing the complications
of the quiet Sun magnetic fields also has
limitations.  Equations~(\ref{pdfIV}) and 
(\ref{pdfIVb}) imply a one-to-one relationship between 
the magnetic field and the thermodynamic structure of the
atmosphere. Although a coupling between magnetic field
and thermodynamics is to be expected, 
the real relationship should have a significant 
scatter \citep[e.g.][]{vog03b}.
Magnetic field variations along the LOS
are restricted, and  the overlapping along the LOS of 
structures with different magnetic fields is ignored.
Our PDF approach is only a first approximation to the
problem.

\begin{figure}
\plotone{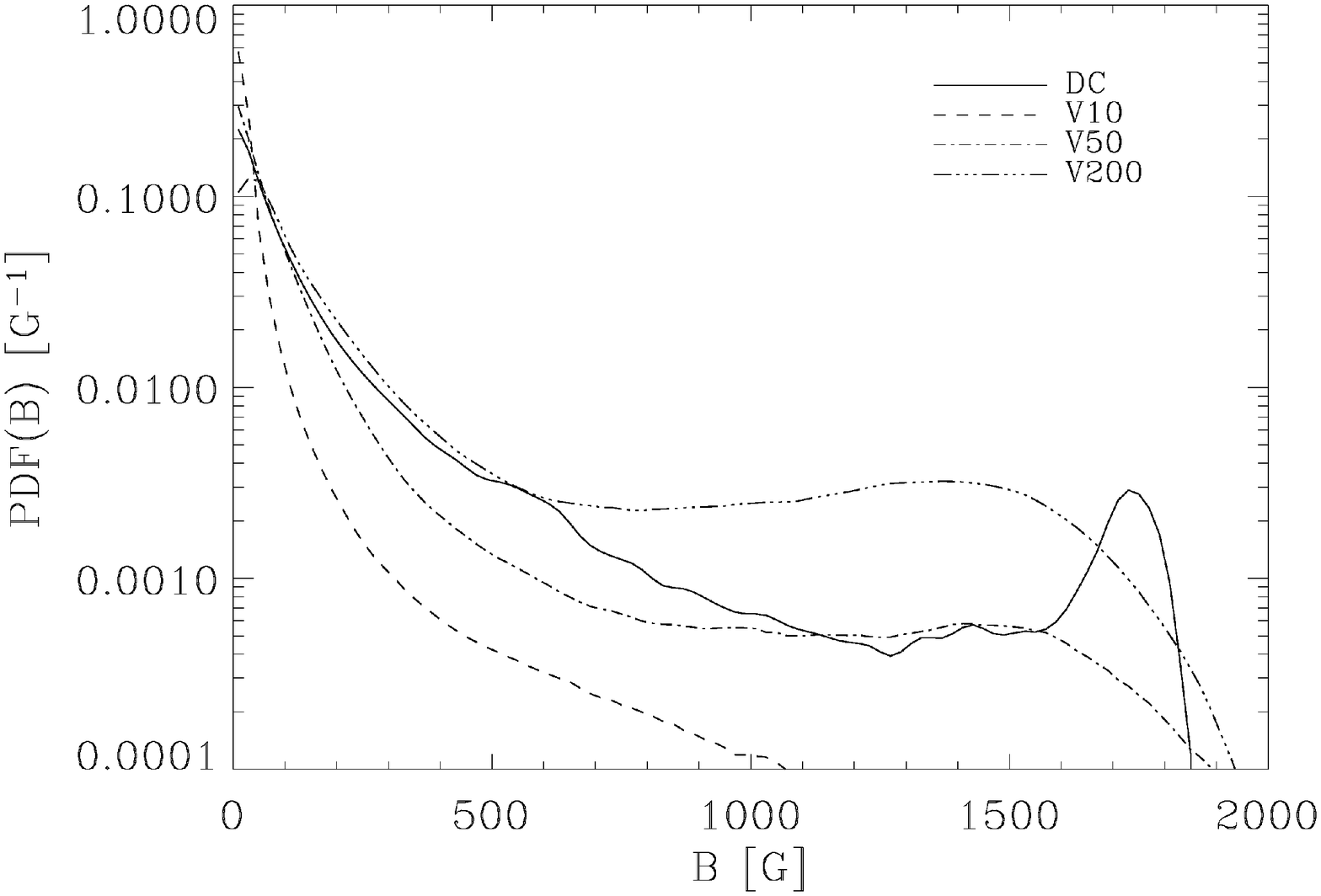}
\caption{Magnetic field strength PDFs used for multi-component syntheses. 
The one labelled DC is from \citet{dom06} while the V10, V50 and V200 are from 
\citet{vog03b}. See text for details.}
\label{pdfs}
\end{figure}

\begin{figure}
\plotone{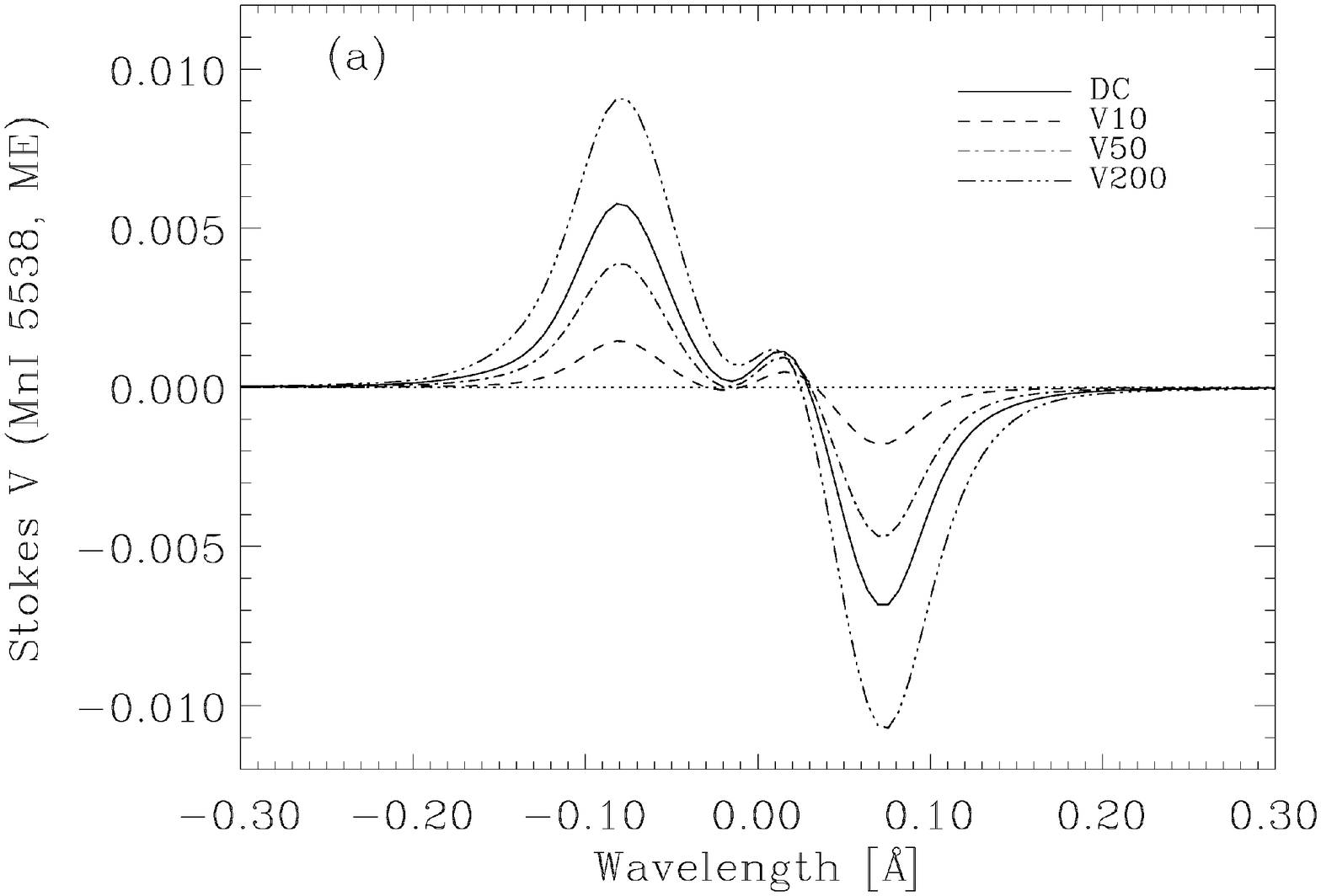}
\plotone{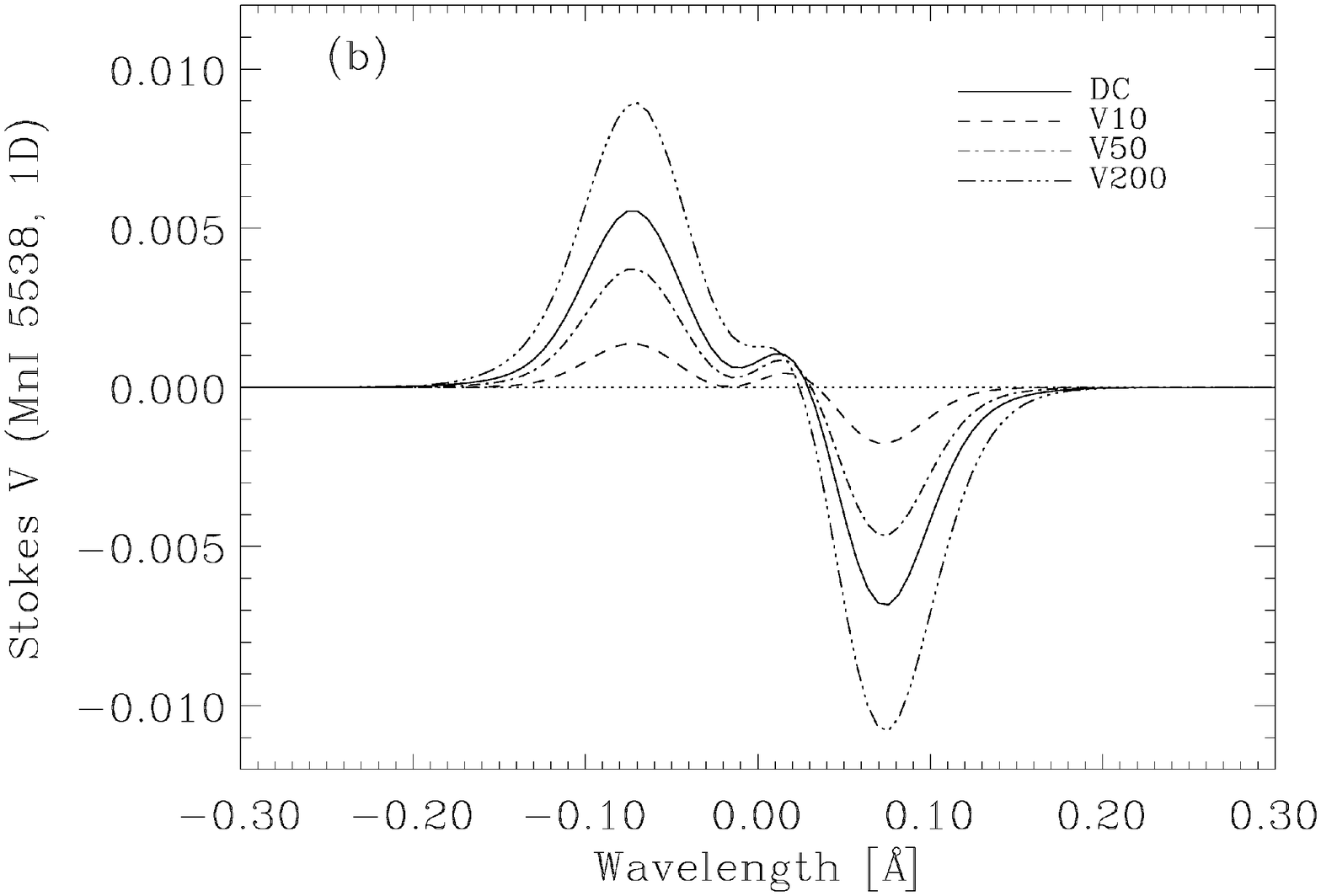}
\plotone{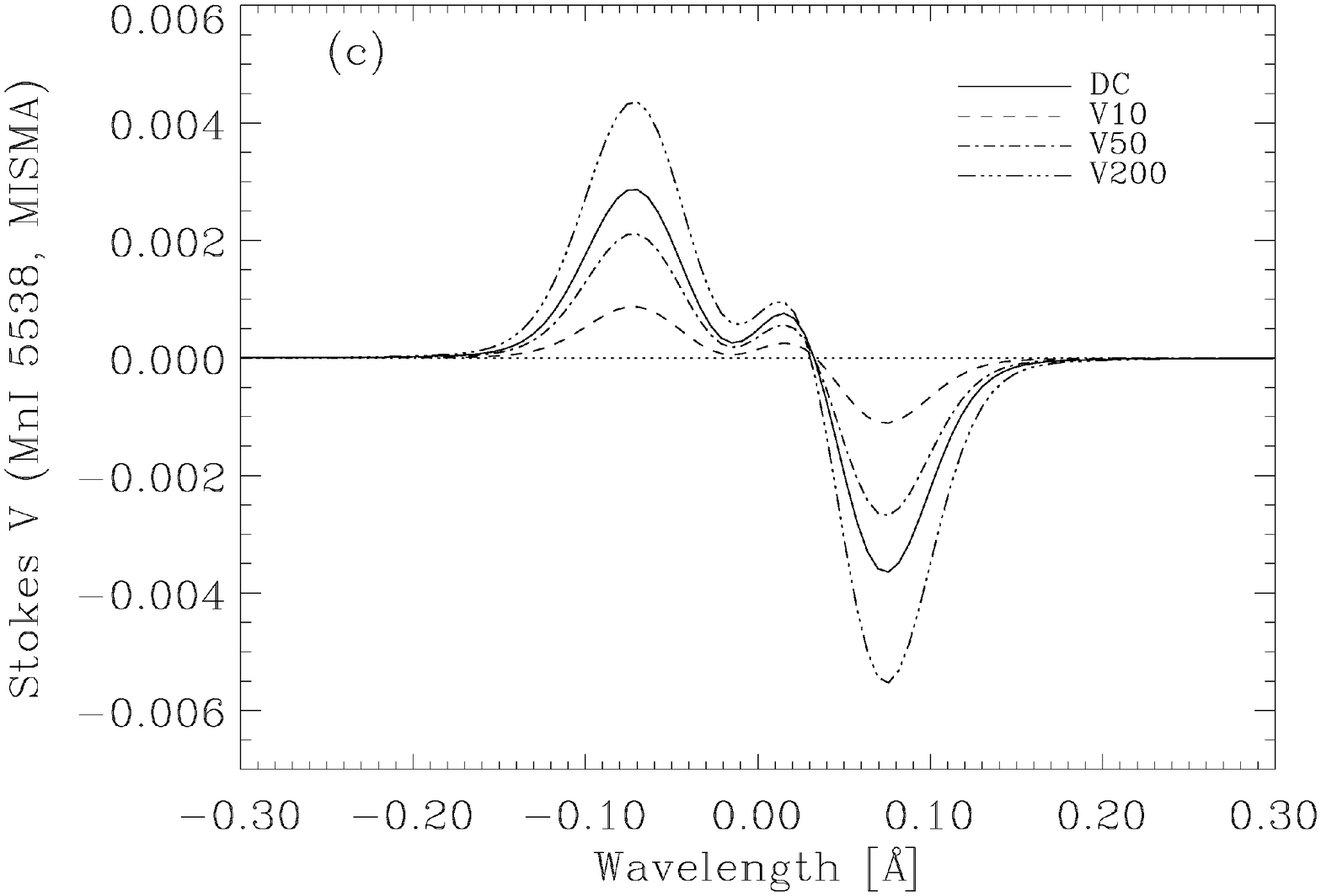}
\caption{Multi-component Stokes~$V$ profiles synthesized using  ME atmospheres (a),
1D atmospheres (b), and MISMAs (c). The labels and the
lines in the insets refer to the PDFs in Fig.~\ref{pdfs}.
The profiles are normalized to the quiet Sun continuum intensity.
}
\label{vsme}
\end{figure}

The four PDFs in Fig.~\ref{pdfs} are used for synthesis. They correspond 
to one of the
semi-empirical quiet Sun PDF obtained by \citet{dom06} 
(labelled DC), plus the PDFs of
the three magneto-convection numerical simulations by \citet{vog03b},
representing magnetic fluxes of  10~G, 50~G and 200~G (V10, V50 and V200, 
respectively).
When ME atmospheres, 1D atmospheres, and model MISMAs 
are combined according to these PDFs, one
obtains the Stokes~$V$ profiles shown in Fig.~\ref{vsme}. 
In this case the magnetic field inclination is constant and set to $60^\circ$,
i.e., $\langle\cos\theta\rangle=1/2$.
The corresponding Stokes~$I$ profiles for the ME and the 1D cases are 
practically identical to those in Fig.~\ref{si}.  
The MISMA synthesis has a continuum intensity 5\%  lower
than that of the quiet Sun, since the thermodynamic structure of 
the model MISMA used for synthesis is cooler than the typical quiet Sun 
model atmospheres \citep[see Fig.11 in][]{san00}. 
 
\begin{figure}
\plotone{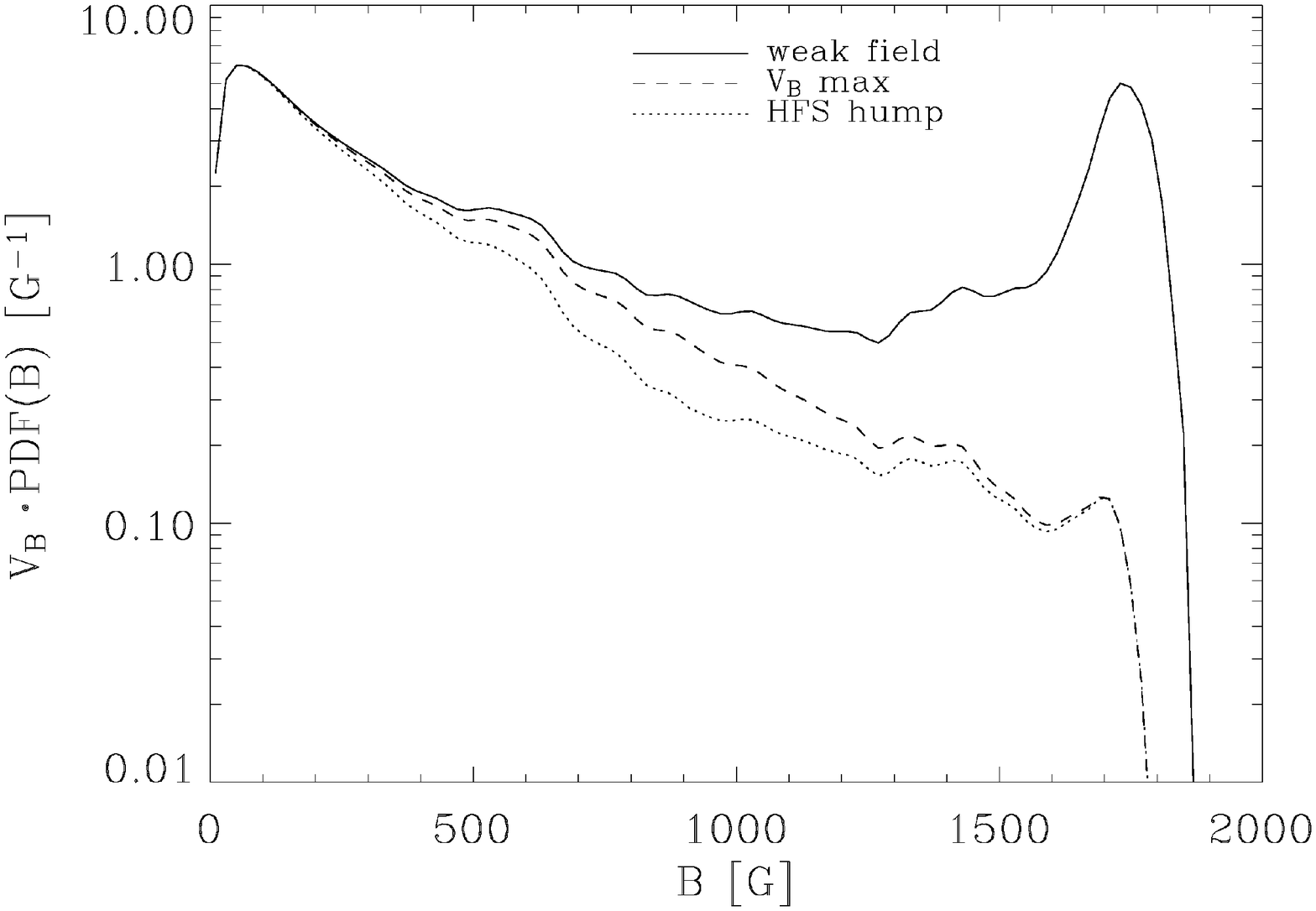}
\caption{Relative contribution to the average Stokes~$V$ profile
of a multi-component atmosphere characterized by 
the PDF labelled as DC in Fig.~\ref{pdfs}.
The integral of the function represented in the plot
is proportional to the Stokes~$V$ signal at
a particular wavelength. The three curves correspond 
to any wavelength under the weak field approximation (the solid line), 
the wavelength where the average Stokes~$V$ 
signals is maximum (the dashed line), and the wavelengths where
the HFS hump shows up (the dotted line). 
The units of the ordinates are relative since the curves have been scaled to the 
solid line when $B\rightarrow 0$.
} 
\label{phi}
\end{figure}

Note that the Stokes~$V$ profiles in Figs.~\ref{vsme} always 
show HFS humps. This result warns us against simplistic 
interpretations of the presence of HFS features as the
unequivocal signature of an atmosphere dominated by hG magnetic 
fields. Despite the fact that most of the atmospheric volume
has very low field strengths, the magnetic flux and the 
magnetic energy of some of the PDFs used for synthesis are
dominated by the tail of kG fields. The extreme  case corresponds
to  V200 in Fig.~\ref{pdfs}, where 60\% of the
magnetic flux and 87\% of the 
energy\footnote{Following \citet{san04} and \citet{dom06},
the unsigned magnetic flux and the magnetic energy are computed
as the two first moments of the PDF.}
is in field strengths larger
than 700~G, which is the field strength where the HFS hump
disappears according to the ME syntheses.
Even in this case the Stokes~$V$ profiles present 
clear HFS humps (see Fig.~\ref{vsme}, the triple 
dotted-dashed lines).
In order to understand this behavior one has to realize 
that the various magnetic field strengths existing in the 
atmosphere do not contribute to the {\em observable} signal 
in proportion
to their magnetic flux, as expected if the weak magnetic
field approximation would be satisfied (equation~[\ref{wfa}]). 
The strong kG fields are always under-represented (\S~\ref{scs}), 
and this bias is particularly severe at line center where 
the HFS hump shows up. The effect is shown
in Fig.~\ref{phi} which contains the 
integrand of equation~(\ref{pdfIVb}) for one
of the PDFs in Fig.~\ref{pdfs} (DC). The integrand is given
in three cases, (a)
assuming the weak field approximation to hold (the solid line),
(b) at the wavelength where the average Stokes~$V$
is maximum (the dashed line), and  (c) at the line core
(the dotted line). 
The individual spectra have been synthesized in
model MISMAs, 
but the qualitative behavior is common to all 
other model atmospheres. According to Fig.~\ref{phi} the
contribution of the kG fields to the HFS hump of the average Stokes~$V$ signal 
is some ten times smaller than that predicted by the weak field 
approximation.  The relative contribution
of the various field strengths to the hump is worked out as part
of \S~\ref{proxy}. 

\subsection{On the diagnostic provided by the HFS hump}
\label{proxy}

\begin{figure}
\plotone{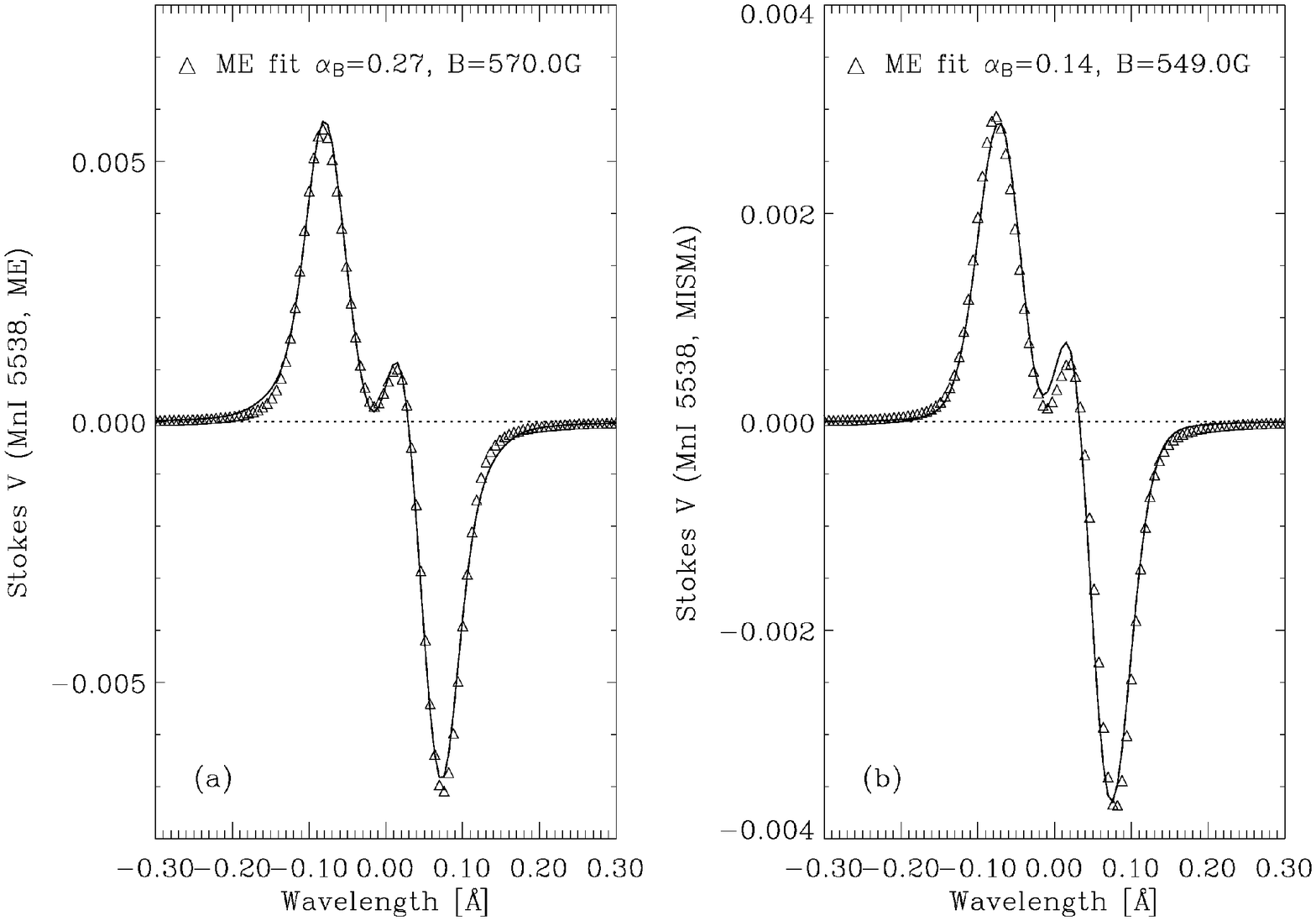}
\caption{Least squares fits of Stokes~$V$ profiles 
synthesized in multi-component atmospheres (the solid lines) 
using single-component ME atmospheres (the symbols). 
(a) The multi-component Stokes~$V$ profile is synthesized in 
ME atmospheres. 
(b)  The multi-component  Stokes~$V$ profile is synthesized in 
model MISMAs. 
The values for $\alpha_B$ and $B$
retrieved from the fits are included in the figures. The two
synthetic profiles correspond to the same PDF (DC).
The profiles are normalized to the quiet Sun continuum intensity.
} 
\label{fit}
\end{figure}

The observations of \mni{5538} analyzed in the literature
employ single component ME atmospheres 
(see \S~\ref{intro}). The observed Stokes~$V$ 
profiles are fitted with synthetic profiles to obtain mean 
magnetic field strengths. In order to explore the diagnostic 
content of such inversion technique when applied to complex magnetic
atmospheres, we developed a simple 
least squares minimization procedure to fit single-component
ME profiles $V_B$ to the synthetic multi-component $V$. 
The procedure fits
\begin{equation}
V= \alpha_B\ V_B,
\end{equation}
where the factor $\alpha_B$ and the field strength $B$ are the 
only two unknowns. 
Figure~\ref{fit} shows the single-component ME fit of
the multi-component Stokes~$V$ produced by the DC PDF.
In the case that the syntheses are based on ME atmospheres 
(Fig.~\ref{fit}a), the match is excellent. If the
synthesis is based on model MISMAs, the fit worsens a bit but 
still the agreement is fair (Fig.~\ref{fit}b). In both
cases the field strength resulting from the ME {\em inversion}
is about 550~G. We have unsuccessfully tried to link this 
average  magnetic field strength with a particular 
property of the underlying  PDF (the solid line in Fig.~\ref{pdfs}). 
It is neither the unsigned flux ($\sim$150~G) nor the 
most probable field strength ($\sim$13~G). Figure~\ref{phi},
the solid line, shows $B P(B)$, i.e., the unsigned flux 
per unit of magnetic field strength. It peaks at 50~G and 1700~G
which, again, do not seem to  bear 
any obvious relationship 
with the average ME  magnetic field strength ($\sim$550~G). 

In order to understand the relationship between the HFS hump 
and the magnetic fields existing in the multi-component
atmosphere, we adopt a strategy avoiding ME fits.
The presence or absence of a hump is certainly associated with the
balance between weak fields and strong fields, but it is
unclear how the hump is related with the fraction of atmosphere filled
by weak and strong fields. One can quantify the
presence of the hump using the derivative of the
Stokes~$V$ profile at a wavelength $\lambda_c$ slightly bluewards
of the HFS maximum (i.e., centered in linear growth that
ends up in the hump; see Fig~\ref{svb}a where $\lambda_c\simeq$0). 
Then the sign of 
\begin{equation}
{\frac{dV}{d\lambda}}\Big\vert_{\lambda_c},
\label{weak}
\end{equation}
indicates the presence ($>0$) or absence ($< 0$) of a
hump\footnote{This criterion assumes positive magnetic polarity, implying a positive Stokes~$V$ 
blue lobe. The same rule applies to negative polarity if one
reverses the sense of the inequalities.}. 
According to equation~(\ref{pdfIVb}),
\begin{equation}
{\frac{dV}{d\lambda}}\Big\vert_{\lambda_c}=
\int_{0}^{B_{\rm max}}{\langle\cos\theta\rangle\,P(B)\,{\frac{dV_B}{d\lambda}}\Big\vert_{\lambda_c}\,dB},
\label{intder}
\end{equation}
so the slope of the mean Stokes~$V$ is just a weighted mean 
of the slopes that characterize each field strength. Consequently,
$dV_B/d\lambda\vert_{\lambda_c}$ yields the contribution of each 
field strength to the sign of the mean
profile or, in other words, it yields 
the contribution of each field strength to the presence or
absence of a hump in the average profile.

\begin{figure}
\plotone{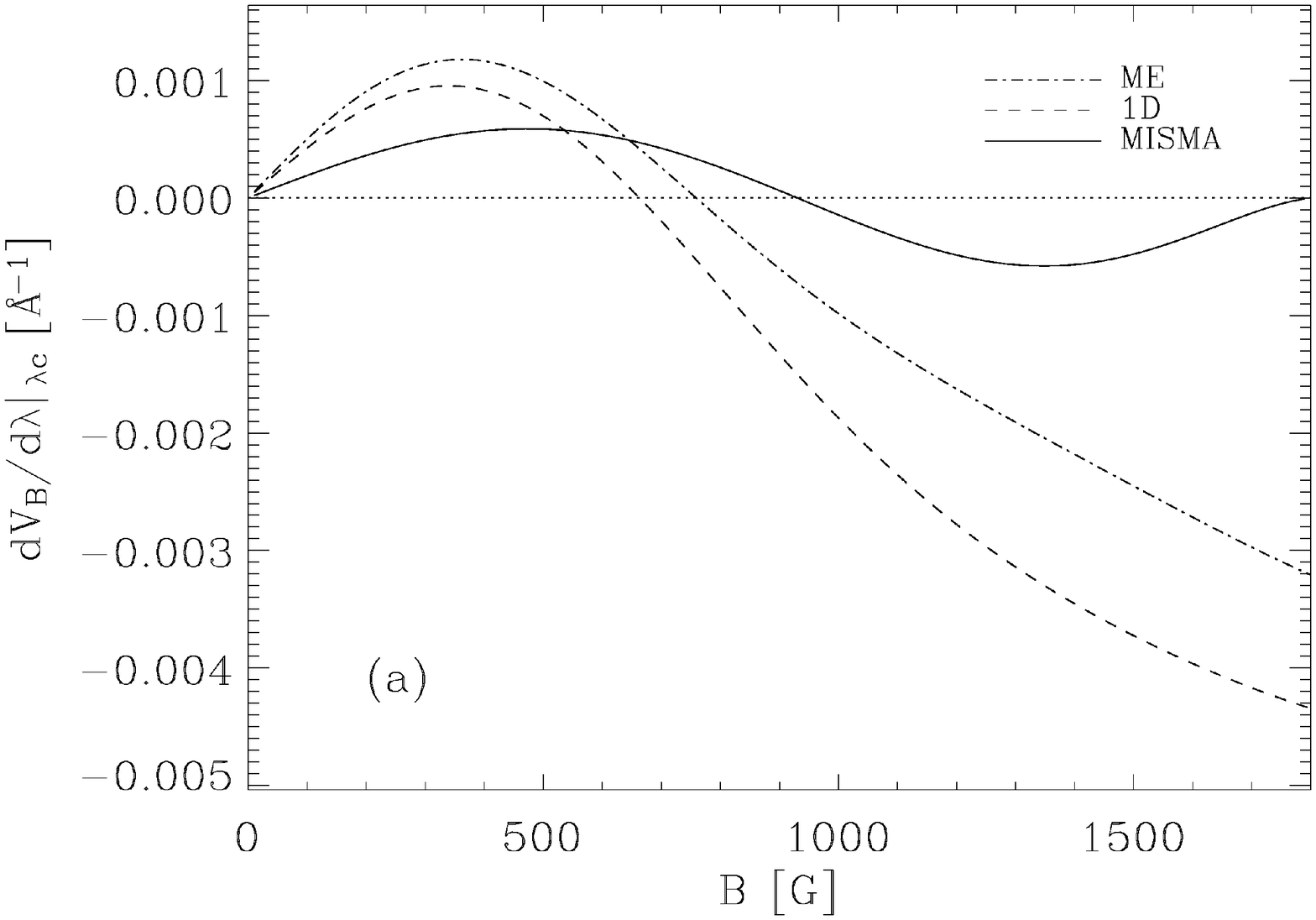}
\plotone{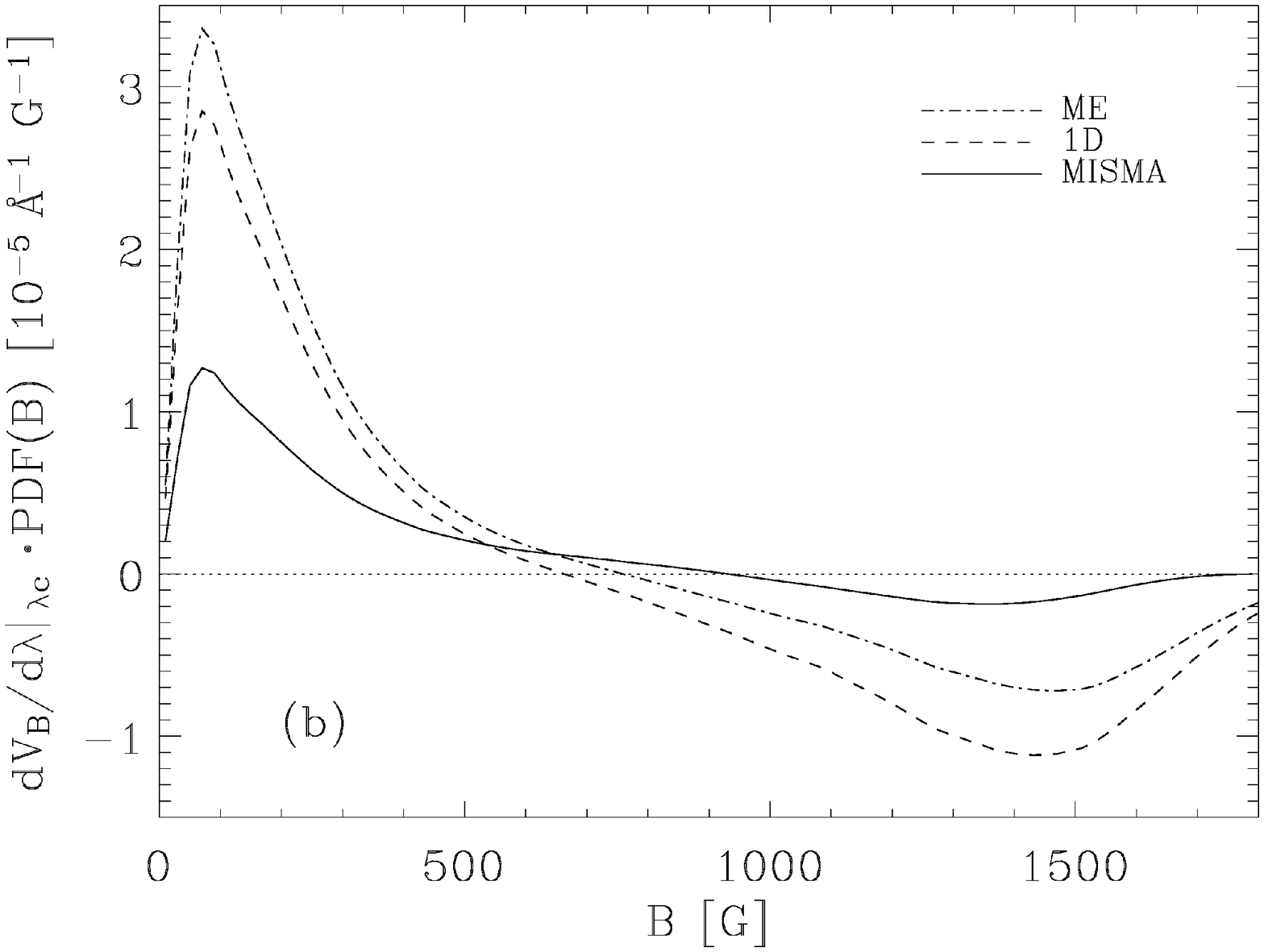}
\caption{(a) Sensitivity of the HFS hump to the 
distribution of magnetic field strengths in the resolution element. 
The hump appears or goes away depending on the integral of   
$dV_B/d\lambda\vert_{\lambda_c}$ weighted with the PDF. 
The three curves correspond to the three model atmospheres used
for multi-component synthesis (see the inset).
(b) The derivative shown in  (a) weighted with the 
V200 PDF. 
The integral
is positive meaning that the resulting Stokes~$V$ profiles 
have HFS humps; see the 
triple dotted-dashed
lines in Fig.~\ref{vsme}.
}
\label{der}
\end{figure}

We have applied equation~(\ref{intder}) to understand the 
HFS humps of the synthetic profiles
in Fig.~\ref{vsme}. 
Figure~\ref{der}a shows the
Stokes~$V$
derivative at line core ($\lambda_c=0$)
for the ME, the 1D and the MISMA model atmospheres. 
The magnetic field strength where this 
derivative is zero indicates when $V_B$ changes from having
to laking of a hump. As the figure shows, this transition 
field strength depends on the model atmosphere.  
The positive and negative lobes of the curves in 
Fig.~\ref{der}a 
have similar amplitudes,  which indicates that for the average Stokes profile 
to lack of a hump, the filling factor of the kG fields  in the atmosphere 
has to be similar to the filling factor in weak fields. 
The requirement is not fulfilled
by the PDFs in Fig.~\ref{pdfs},
whose filling factors are dominated by weak fields. Such 
superiority 
explains why all profiles in 
Fig.~\ref{vsme} show HFS humps. 
Equation~(\ref{intder}) also explains why the magnetic flux in the form 
of kG structures has to exceed by far the flux in weak fields
for the HFS hump to disappear.
The HFS hump goes away when 
weak fields and strong fields have similar filling factors,
but then the kG fields dominate the 
flux since the (unsigned) magnetic flux scales
as the filling factor times the  field strength. 
For example, consider a resolution element 
filled half-and-half with structures of 100~G and 1~kG. 
The HFS hump is not present, but the 
magnetic flux in kG fields is ten times the weak field
flux.

Among the three
curves in Fig.~\ref{der}a, the one 
corresponding to MISMAs reflects the solar
conditions best, since it includes the dimming of the line 
with increasing field strength.  This curve has positive and 
negative lobes of approximately the same area, therefore, 
the presence-absence of HFS humps in the 
\mni{5538} Stokes~$V$
profiles is controlled by the fraction 
of photospheric atmosphere with field 
strengths smaller-larger than some 1~kG (i.e., the zero 
crossing point of the curve). Interpreting real observations
with this rule-of-thumb
requires disregarding
the atmospheric volume having very weak magnetic 
fields, which produce no polarization and therefore 
do not contribute to the observed signal
(see Fig~\ref{der}b).

\section{Mixed polarities and unresolved velocities}\label{comp_obs}

\begin{figure}
\plotone{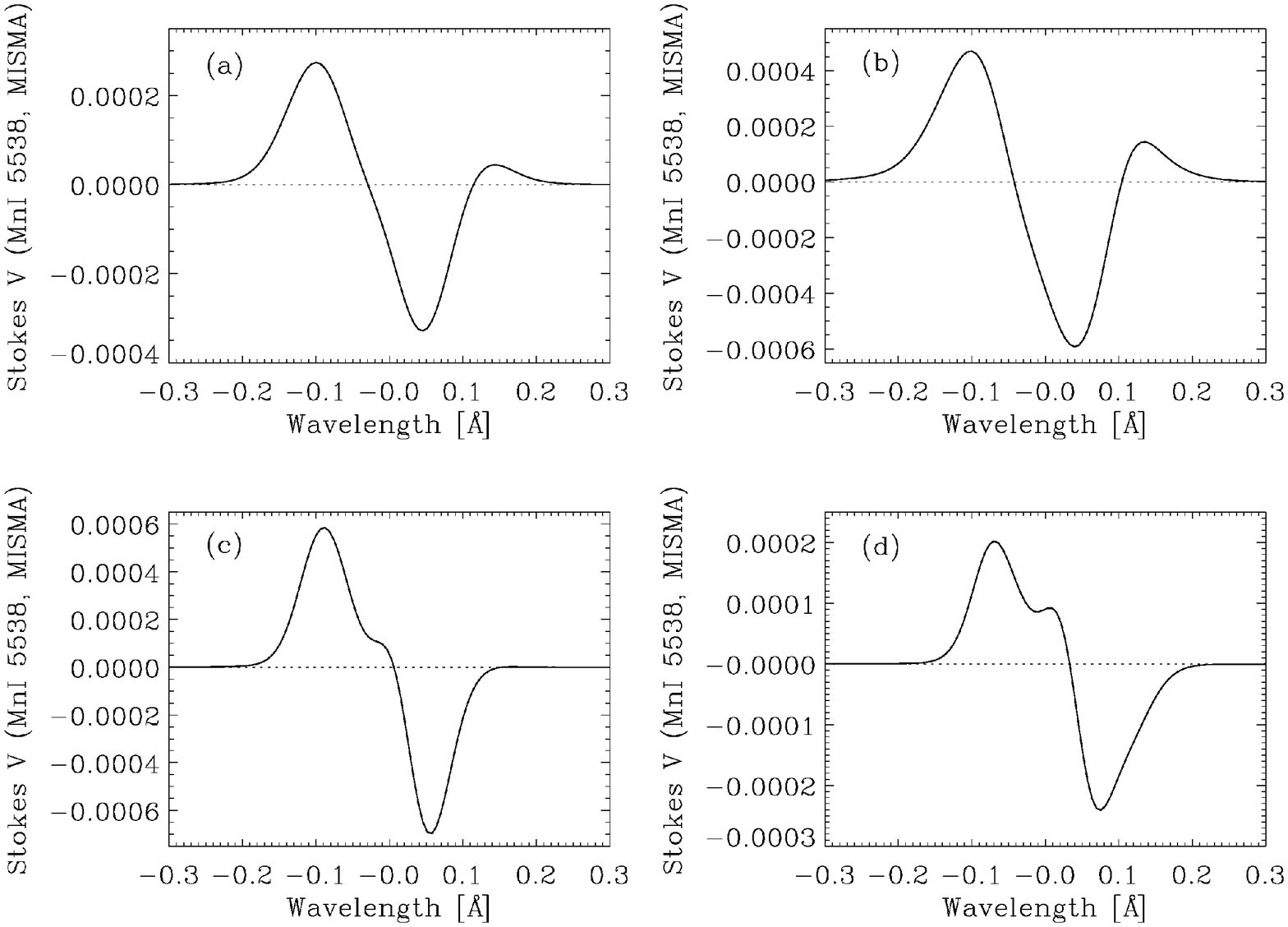}
\caption{
Examples of Stokes~$V$ profiles synthesized in semi-empirical
model MISMAs having unresolved mixed polarities correlated 
with velocities. They correspond to  different classes in \citet{san00}
-- (a) class 6, 
(b) class 9, and (c) class 7. (d) Offspring of 
the class 7 
model atmosphere where the two polarities have the 
same sign and the velocity difference has been increased.
The profiles are normalized to the quiet Sun continuum intensity.
} 
\label{misma0}
\end{figure}

We expect the quiet Sun resolution elements to contain mixed 
polarities \citep{san00,soc02}, which leads to a reduction 
of  the Stokes~$V$ signals. In the multi-component approximation
parlance (equation~[\ref{pdfIVb}]), having
unresolved mixed polarities implies,
\begin{equation}
\langle\cos\theta\rangle \ll \langle|\cos\theta|\rangle, 
\end{equation}
since positive and negative $\cos\theta$ tend to cancel when
carrying out the averages. If 
the positive polarity ($\cos\theta > 0$) and the negative
polarity ($\cos\theta < 0$) have the same distribution of 
inclinations but they fill different
fractions of resolution element, then
\begin{equation}
\langle\cos\theta\rangle=  \langle|\cos\theta|\rangle\, (2f-1), 
\label{mixpol}
\end{equation}
with $f$ the fraction filled by the magnetic fields of positive 
polarity. If $f$ does not depend on the magnetic 
field strength, the effect of mixed polarities is only a 
global scaling of the Stokes~$V$ profile with no effect 
on Stokes~$I$ (equations~[\ref{pdfIV}], [\ref{pdfIVb}], 
and [\ref{mixpol}]).
However,  the above
view of the effect produced by unresolved mixed polarities
is too simplistic. It implicitly neglects the existence of unresolved 
velocities in the resolution element coupled with changes of polarity. 
When the sense of the velocity and the polarity are correlated,
the Stokes~$V$ of opposite polarities 
are Doppler shifted. They
do not perfectly cancel,
leaving observable residuals. Such coupling 
has been observed in Fe lines,
and it produces very asymmetric 
Stokes~$V$ profiles \citep{san00}. In order to illustrate the
effect, Fig.~\ref{misma0} shows Stokes~$V$
profiles of \mni{5538} synthesized in empirical model MISMAs
with mixed polarities and different velocities \citep[][]{san00}.

\begin{figure}
\plotone{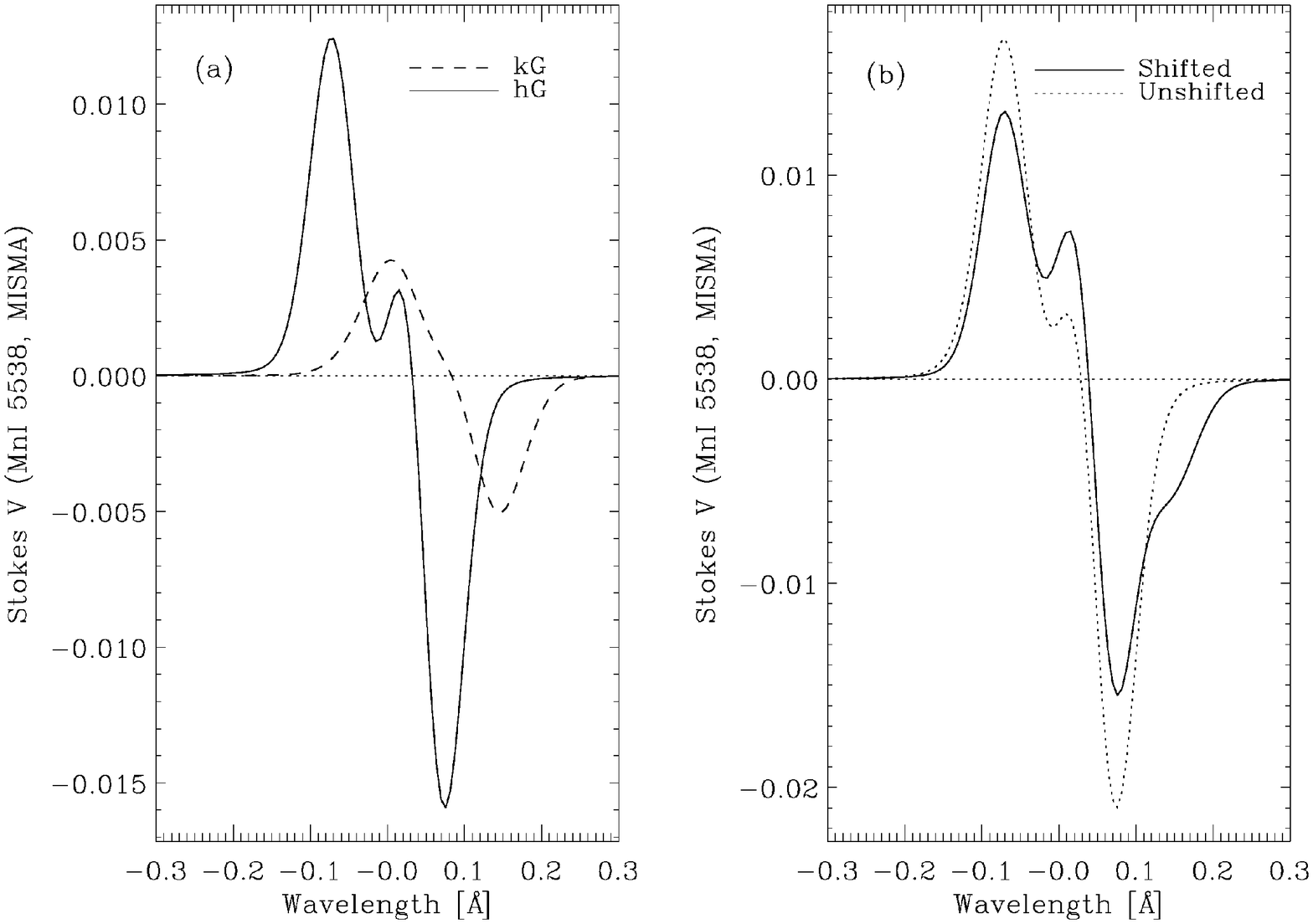}
\caption{
	Two component Stokes~$V$ synthesis of weak and strong fields 
	having different Doppler shifts. The strong fields are 
	blueshifted with respect to the weak fields -- see (a). 
	The 
	sum
	of the profiles in (a) leads to the profiles in
	(b), the solid line. If the relative Doppler shift is not
	considered, then the average Stokes~$V$ corresponds
	to the dotted line in (b).
The profiles are normalized to the quiet Sun continuum intensity.
}
\label{twocomp}
\end{figure}

Velocities and other magnetic 
quantities
 are also expected to be
coupled. For example, strong kG 
fields appear in intergranular lanes with downflows, 
whereas weak fields prefer granules with upflows 
\citep[e.g., ][]{cat99a,soc04}. This combination of Doppler 
shifts and magnetic field strengths renders profiles like the two
component average shown in Fig.~\ref{twocomp}b, the solid
line. It corresponds to the  sum
of the profiles 
in Fig.~\ref{twocomp}a, which are shifted by some
3.5 km~s$^{-1}$. The HFS hump now appears 
elevated from the zero level, and the Stokes~$V$ blue wing is 
broadened with a conspicuous double aborption structure. 
These two features are common among the observed
Stokes~$V$ profiles of \mni{5538} \citep[][ Figs. 8 and 9]{lop06}.
The same features appear in the MISMA synthesis
of Fig.~\ref{misma0}d. 

In short, mixed polarities and velocity gradients are
properties of the real quiet Sun magnetic fields 
that affect the observed polarization in a significant way.
They are not easy to deal with but 
must be taken into account for a realistic interpretation
of the observations.

%
\section{Synthesis of other \mnis{} lines}
\label{other}
The purpose of synthesizing \mnis{} lines different from
\mni{5538} is threefold. First, it confirms that our
LTE syntheses reproduce all kinds of observed
HFS patterns. Second, it offers a chance to explore whether 
other \mnis{} lines show a sensitivity to magnetic fields 
complementary to \mni{5538}.
Finally, it proves that the uncertainty of the HFS 
constants do not limit the use of Mn~{\sc i} lines for 
magnetometry.

\begin{figure}
\plotone{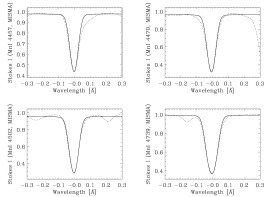}
\caption{\mnis{} lines without obvious HFS features in the observed solar spectrum. 
The solid lines show our syntheses whereas the dotted lines correspond
to the solar atlas \citep{nec99}. Wavelengths,  given in \AA ,
are refereed to the laboratory wavelength of the lines.
Each line is identified by the wavelength in the label
of the ordinate axis.
The profiles are normalized to the quiet Sun continuum intensity.
}
\label{nohfs}
\end{figure}

\begin{figure}
\plotone{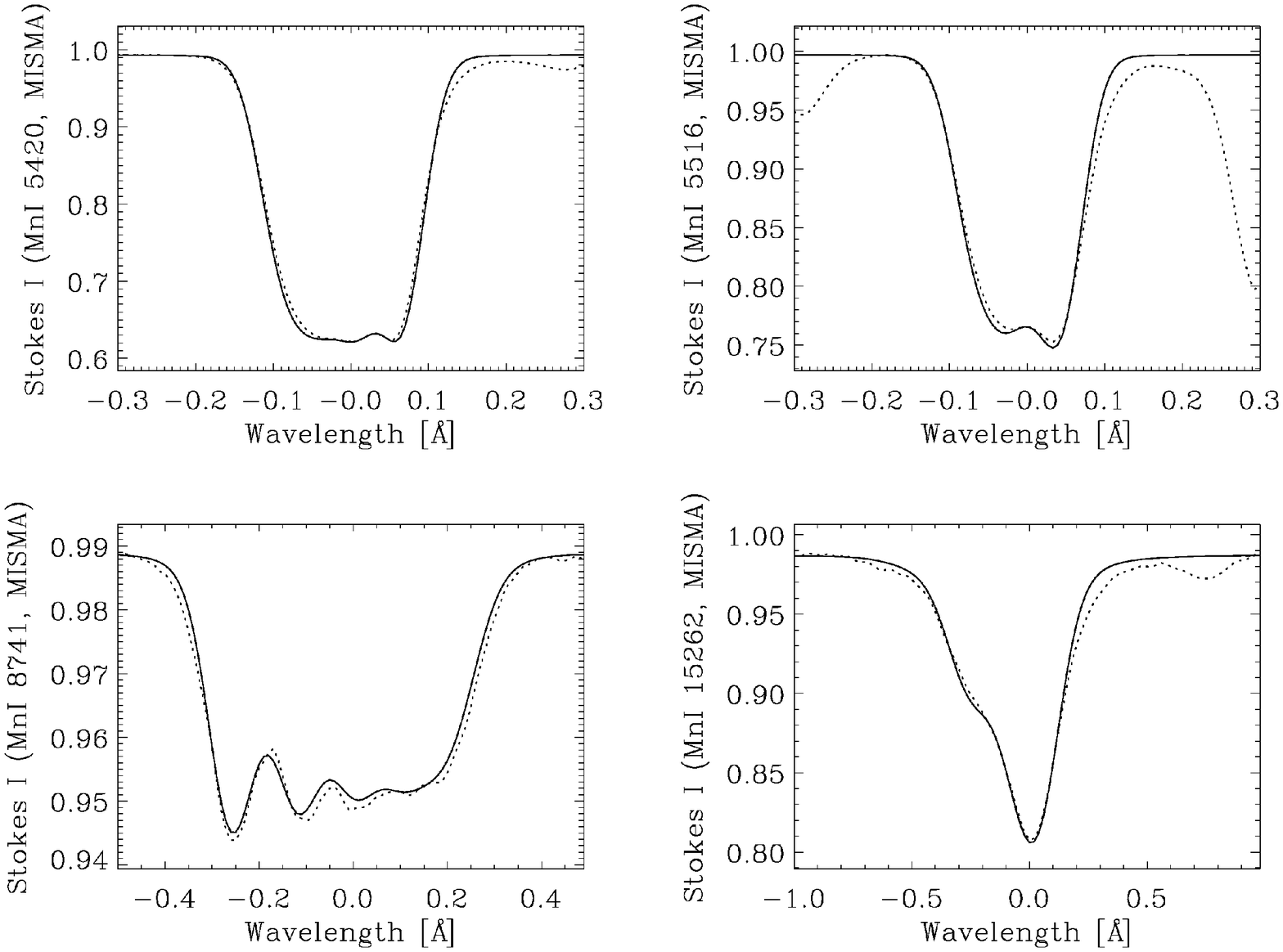}
\caption{\mnis{} lines with  HFS features. 
The solid lines correspond to our syntheses whereas the dotted lines show
the solar atlas \citep{nec99}. 
Wavelengths,  given in \AA ,
are refereed to the laboratory wavelength of the lines.
Each line is identified by the wavelength in the label
of the ordinate axis.
The profiles are normalized to the quiet Sun continuum intensity.
}
\label{sihfs}
\end{figure}

All the syntheses in this section employ
the non-magnetic quiet Sun MISMA that reproduces the
\mni{5538} profile in Fig.~\ref{si}. Varying 
the macroturbulence and, in some cases also
tuning the oscillator strength, we managed to fit the
observed Stokes~$I$ profiles of all
the lines we tried\footnote{The pattern of \mni{5407}
puzzled us for some time. As suggested
by R. Manso, it turned out to be a blend of the \mnis{} line with two fairly
strong \feis{} lines that  remain unidentified in the classical list 
by \citet{moo66}, and  which do not
appear in the NIST Atomic Spectra Database. We found them with 
approximately the observed strengths among the 
list of lines by Kurucz at {\tt http://kurucz.harvard.edu/linelists.html}.}. 
A total of eight \mnis{} lines 
reproduced by the code are shown in Figs.~\ref{nohfs} and \ref{sihfs}.
They are split in two sets; Fig.~\ref{nohfs} includes lines without 
obvious HFS features in the observed solar spectrum, 
whereas Fig.~\ref{sihfs} shows lines with 
varied and assorted HFS.
The agreement between the synthetic profiles and the solar atlas is good, an
impression reinforced by the fact that the HFS constants determining the 
HFS patterns 
have not been tuned, but they come directly from the literature (Table~\ref{param}).

The dependence of the various lines on magnetic field strengths has been 
studied by synthesizing them in the magnetic quiet Sun model MISMA 
used in \S~\ref{mcs}, increasing the field strength from $\sim 100~$G to 
$\sim 2000~$G. All the lines share a common behavior -- they weaken with 
increasing field strength, and the HFS features tend to disappear. 
This behavior is also shared by the near IR line \mni{15262} recently
pointed out by \citet{ase07} -- Fig.~\ref{sihfs}, bottom right.
On top of this general trend, and depending on the actual HFS constants,
the Stokes~$V$ profiles of different lines present specific
features. The line \mni{5516} behaves like \mni{5538}, with a single HFS reversal
at the line core. 
Then it is possible to relate the presence or absence of 
the hump with the distribution of field
strengths in the atmosphere, exactly as the analysis carried out
in \S\ref{proxy}. The patterns of \mni{5420} and \mni{8741} are more complicated, 
and we have not been able to agree on a particular feature of the Stokes~$V$ 
profiles that can be treated as in \S\ref{proxy}.

\section{Discusion and conclusions}\label{conclusions}

The Zeeman pattern of the Mn~{\sc i} lines depends on
hyperfine structure (HFS), which confers them a 
sensitivity to hG magnetic field strengths different 
from the  lines traditionally used in solar 
magnetometry. This peculiarity has been used to measure 
magnetic field strengths in quiet Sun regions 
(see \S~\ref{intro}). The methods applied
so far assume the magnetic field strength to be constant
in the resolution element, an approximation driven
by feasibility rather than based on physical or 
observational arguments. Actually, it is not a
good approximation since the magnetic fields 
of the quiet Sun are expected to vary on very small spatial
scales, with field strengths spanning from zero to 2kG. 
Under these extreme conditions, all diagnostic techniques 
employed in quiet Sun magnetometry are strongly biased
towards a particular range of field strengths,
and the Mn~{\sc i} signals are not expected to be 
the exception. Consequently, the  diagnostic content of 
the Mn~{\sc i} lines cannot be fully exploited unless their 
biases are properly understood. Such task is undertaken
in the paper by exploring the response
of Mn~{\sc i} lines in a number of realistic quiet 
Sun scenarios. 

Three complementary LTE synthesis codes 
have been written and tested (ME, 1D and MISMA; 
\S~\ref{code}). They provide a number of relevant results,
the first one being the ability to reproduce 
all observed Mn~{\sc i} HFS patterns. We reproduce
the observed unpolarized line profile of nine assorted lines 
corresponding to  all HFS sensitivities (\S~\ref{other}).
The study is focused on the line most often used
in magnetometry,  \mni{5538}, however its behavior
should be representative
of the other lines.
According to the weak magnetic field approximation, the
Stokes $V$ signals scale with the longitudinal
component of the magnetic field, i.e.,
the magnetic field strength times the cosine of the 
magnetic field inclination with respect to the LOS. We verify that
the scaling on cosine holds very tightly. The 
scaling on the magnetic field strength, however,
breaks down soon  (when $B \ge  400$~G for 
\mni{5538}). Even for ME  atmospheres, the weak field approximation 
predicts a  Stokes~$V$ signal twice as large as the 
synthetic one for $B\simeq$1.5~kG. 
When the expected coupling between the thermodynamic 
conditions and the magnetic field strengths is 
taken into account, the dimming of the kG Stokes~$V$ 
signals can be as large as two orders of magnitude  
(see Figs.~\ref{maxv} and \ref{phi}).
The dimming of the polarization signals formed in 
kG magnetic concentrations affects all Mn~{\sc i} lines
(\S~\ref{scs}),  and it has significant observational implications.

If the resolution elements contains both weak and strong fields,
then the kG fields tend to be under-represented in the
average profile. We have modelled the bias assuming a 
multi component atmosphere, where the synthetic signals 
are weighted means of the Stokes profiles 
corresponding to each single field strength. The weight
is given by the fraction of atmosphere filled by 
each field strength, i.e., by the magnetic
field strength probability density function PDF
(equations~[\ref{pdfIV}] and [\ref{pdfIVb}]). 
According to our modeling, even when the (unsigned)
magnetic flux and the magnetic energy are dominated
by kG magnetic fields, the Stokes~$V$ profile of \mni{5538}
can show HFS reversal at line core
characteristic of hG fields
(Fig.~\ref{vsme}). A pure morphological inspection of the
Stokes profiles  does not suffice to infer which is the 
dominant magnetic field strength in the resolution element. 
For the HFS hump of \mni{5538} to disappear the kG filling factor 
has to be larger than the sub-kG filling factor and, consequently,
when the HFS hump disappears the magnetic flux and 
magnetic energy of the atmosphere are completely
dominated by kG fields (\S~\ref{proxy}).

%

Detecting Stokes~$V$ profiles with HFS features 
indicates the presence of hG fields in the
resolution element. However, this sole observation
does not tell whether the hG field strengths dominate.
There seem to be two extreme alternatives to exploit the
diagnostic potential of these lines. First, 
improving the spatial resolution of the 
observations to a point where the 
quiet Sun magnetic structures can be regarded as 
spatially resolved. Unfortunately, this possibility does 
not seem to feasible at present. Realistic simulations of 
magneto-convection indicate that quiet Sun magnetic 
fields are uniform only at the diffusive length scales
\citep[e.g.][]{cat99a,vog07}, 
which are of the order of a few  km in the photosphere
\citep[e.g.][]{sch86}. These scales are very far from the 
angular resolution of the present measurements, 
and even much smaller than the length-scale for the radiative 
transfer average along the LOS \citep{san96}. 
We prefer the alternative possibility, namely, 
developing inversion techniques where complex 
magnetic atmospheres are included into the diagnostics.
Using appropriate constraints, the
number of free parameters required to describe such 
atmospheres can be maintained within reasonable limits
\citep[e.g., the model MISMAs in ][]{san97b}. 
Dealing with unresolved velocities also 
favors
detailled inversion codes (\S~\ref{comp_obs}).
We 
are presently working on 
these improvements needed to develop 
the diagnostic technique pioneered by 
\citeauthor{lop02}.

\acknowledgements
We thank Arturo L\'opez Ariste and
Rafael Manso for helpful discussions
during the development of this 
work.
BV acknowledges a grant from the
{\em  Universit\`a Tor Vergata}
while visiting the IAC.
This work has been partly funded by the  
Spanish Ministry of Education and Science
(project AYA2004-05792). 
%


%
%
%
%
%
%
%
%
%
%
%
%
%

%

\begin{thebibliography}{56}
\expandafter\ifx\csname natexlab\endcsname\relax\def\natexlab#1{#1}\fi

\bibitem[{{Asensio Ramos} {et~al.}(2007){Asensio Ramos}, {Mart{\'{\i}}nez
  Gonz{\'a}lez}, {L{\'o}pez Ariste}, {Trujillo Bueno}, \& {Collados}}]{ase07}
{Asensio Ramos}, A., {Mart{\'{\i}}nez Gonz{\'a}lez}, M.~J., {L{\'o}pez Ariste},
  A., {Trujillo Bueno}, J., \& {Collados}, M. 2007, \apj, 659, 829

\bibitem[{{Bellot Rubio} \& {Collados}(2003)}]{bel03}
{Bellot Rubio}, L.~R., \& {Collados}, M. 2003, \aap, 406, 357

\bibitem[{{Blackwell-Whitehead} {et~al.}(2005){Blackwell-Whitehead},
  {Pickering}, {Pearse}, \& {Nave}}]{Blac05}
{Blackwell-Whitehead}, R.~J., {Pickering}, J.~C., {Pearse}, O., \& {Nave}, G.
  2005, \apjs, 157, 402

\bibitem[{{Bommier} {et~al.}(2005){Bommier}, {Derouich}, {Landi
  Degl'Innocenti}, {Molodij}, \& {Sahal-Br{\' e}chot}}]{bom05}
{Bommier}, V., {Derouich}, M., {Landi Degl'Innocenti}, E., {Molodij}, G., \&
  {Sahal-Br{\' e}chot}, S. 2005, \aap, 432, 295

\bibitem[{{Cattaneo}(1999)}]{cat99a}
{Cattaneo}, F. 1999, \apjl, 515, L39

\bibitem[{{Dom\'\i nguez Cerde\~na} {et~al.}(2003){Dom\'\i nguez Cerde\~na},
  {S\'anchez Almeida}, \& {Kneer}}]{dom03b}
{Dom\'\i nguez Cerde\~na}, I., {S\'anchez Almeida}, J., \& {Kneer}, F. 2003,
  \aap, 407, 741

\bibitem[{{Dom\'\i nguez Cerde\~na} {et~al.}(2006{\natexlab{a}}){Dom\'\i nguez
  Cerde\~na}, {S\'anchez Almeida}, \& {Kneer}}]{dom06}
---. 2006{\natexlab{a}}, \apj, 636, 496

\bibitem[{{Dom\'\i nguez Cerde\~na} {et~al.}(2006{\natexlab{b}}){Dom\'\i nguez
  Cerde\~na}, {S\'anchez Almeida}, \& {Kneer}}]{dom06b}
---. 2006{\natexlab{b}}, \apj, 646, 1421

\bibitem[{{Emonet} \& {Cattaneo}(2001)}]{emo01}
{Emonet}, T., \& {Cattaneo}, F. 2001, \apjl, 560, L197

\bibitem[{{Faurobert} {et~al.}(2001){Faurobert}, {Arnaud}, {Vigneau}, \&
  {Frish}}]{fau01}
{Faurobert}, M., {Arnaud}, J., {Vigneau}, J., \& {Frish}, H. 2001, \aap, 378,
  627

\bibitem[{{Faurobert-Scholl}(1993)}]{fau93}
{Faurobert-Scholl}, M. 1993, \aap, 268, 765

\bibitem[{Fisher \& Peck(1939)}]{FisP39}
Fisher, R.~A., \& Peck, E.~R. 1939, Phys. Rev., 55, 270

\bibitem[{{Khomenko} {et~al.}(2003){Khomenko}, {Collados}, {Solanki}, {Lagg},
  \& {Trujillo-Bueno}}]{kho02}
{Khomenko}, E.~V., {Collados}, M., {Solanki}, S.~K., {Lagg}, A., \&
  {Trujillo-Bueno}, J. 2003, \aap, 408, 1115

\bibitem[{{Landi Degl'Innocenti}(1975)}]{lan75}
{Landi Degl'Innocenti}, E. 1975, \aap, 45, 269

\bibitem[{{Landi Degl'Innocenti}(1976)}]{lan76}
---. 1976, \aaps, 25, 379

\bibitem[{{Landi Degl'Innocenti}(1978)}]{lan78}
---. 1978, \aaps, 33, 157

\bibitem[{{Landi Degl'Innocenti} \& {Landolfi}(2004)}]{lan04}
{Landi Degl'Innocenti}, E., \& {Landolfi}, M. 2004, Astrophysics and Space
  Science Library, Vol. 307, {Polarization in Spectral Lines} (Dordrecht:
  Kluwer)

\bibitem[{{Lef{\`e}bvre} {et~al.}(2003){Lef{\`e}bvre}, {Garnir}, \&
  {Bi{\'e}mont}}]{Lefe03}
{Lef{\`e}bvre}, P.-H., {Garnir}, H.-P., \& {Bi{\'e}mont}, E. 2003, \aap, 404,
  1153

\bibitem[{{Lin} \& {Rimmele}(1999)}]{lin99}
{Lin}, H., \& {Rimmele}, T. 1999, \apj, 514, 448

\bibitem[{{Lites}(2002)}]{lit02}
{Lites}, B.~W. 2002, \apj, 573, 431

\bibitem[{{L{\'o}pez Ariste} {et~al.}(2007){L{\'o}pez Ariste}, {Mart{\'{\i}}nez
  Gonz{\'a}lez}, \& {Ram{\'{\i}}rez V{\'e}lez}}]{lop07}
{L{\'o}pez Ariste}, A., {Mart{\'{\i}}nez Gonz{\'a}lez}, M.~J., \&
  {Ram{\'{\i}}rez V{\'e}lez}, J.~C. 2007, \aap, 464, 351

\bibitem[{{L{\'o}pez Ariste} {et~al.}(2002){L{\'o}pez Ariste}, {Tomczyk}, \&
  {Casini}}]{lop02}
{L{\'o}pez Ariste}, A., {Tomczyk}, S., \& {Casini}, R. 2002, \apj, 580, 519

\bibitem[{{L{\'o}pez Ariste} {et~al.}(2006){L{\'o}pez Ariste}, {Tomczyk}, \&
  {Casini}}]{lop06}
---. 2006, \aap, 454, 663

\bibitem[{{Maltby} {et~al.}(1986){Maltby}, {Avrett}, {Carlsson},
  {Kjeldseth-Moe}, {Kurucz}, \& {Loeser}}]{mal86}
{Maltby}, P., {Avrett}, E.~H., {Carlsson}, M., {Kjeldseth-Moe}, O., {Kurucz},
  R.~L., \& {Loeser}, R. 1986, \apj, 306, 284

\bibitem[{{Margrave}(1972)}]{Marg72}
{Margrave}, Jr., T.~E. 1972, \solphys, 27, 294

\bibitem[{{Mart{\'{\i}}nez Gonz{\'a}lez} {et~al.}(2006){Mart{\'{\i}}nez
  Gonz{\'a}lez}, {Collados}, \& {Ruiz Cobo}}]{mar06}
{Mart{\'{\i}}nez Gonz{\'a}lez}, M.~J., {Collados}, M., \& {Ruiz Cobo}, B. 2006,
  \aap, 456, 1159

\bibitem[{{Moore} {et~al.}(1966){Moore}, {Minnaert}, \& {Houtgast}}]{moo66}
{Moore}, C.~E., {Minnaert}, M. G.~J., \& {Houtgast}, J. 1966, The Solar
  Spectrum from 2935 \AA~ to 8770 \AA~, NBS Mono. 61 (Washington: NBS)

\bibitem[{{Neckel}(1999)}]{nec99}
{Neckel}, H. 1999, \solphys, 184, 421

\bibitem[{{Ralchenko} {et~al.}(2007){Ralchenko}, {Jou}, {Kelleher}, {Kramida},
  {Musgrove}, {Reader}, {Wiese}, \& {Olsen}}]{nist07}
{Ralchenko}, Y., {Jou}, F.-C., {Kelleher}, D., {Kramida}, A., {Musgrove}, A.,
  {Reader}, J., {Wiese}, W., \& {Olsen}, K. 2007, NIST Atomic Spectra Database
  (version 3.1.2) (Gaithersburg, MD: National Institute of Standards and
  Technology), http://physics.nist.gov/asd3

\bibitem[{{S\'anchez Almeida}(1992)}]{san92a}
{S\'anchez Almeida}, J. 1992, \solphys, 137, 1

\bibitem[{{S\'anchez Almeida}(1997)}]{san97b}
---. 1997, \apj, 491, 993

\bibitem[{{S\'anchez Almeida}(1998)}]{san98c}
{S\'anchez Almeida}, J. 1998, in ASP Conf. Ser., Vol. 155, Three-Dimensional
  Structure of Solar Active Regions, ed. C.~E. {Alissandrakis} \&
  B.~{Schmieder} (San Francisco: ASP), 54

\bibitem[{{S\'anchez Almeida}(2000)}]{san00c}
---. 2000, \apj, 544, 1135

\bibitem[{{S\'anchez Almeida}(2004)}]{san04}
{S\'anchez Almeida}, J. 2004, in ASP Conf. Ser., Vol. 325, The Solar-B Mission
  and the Forefront of Solar Physics, ed. T.~{Sakurai} \& T.~{Sekii} (San
  Francisco: ASP), 115, (astro-ph/0404053)

\bibitem[{{S\'anchez Almeida}(2005)}]{san05}
---. 2005, \aap, 438, 727

\bibitem[{{S\'anchez Almeida} {et~al.}(2003){S\'anchez Almeida}, {Emonet}, \&
  {Cattaneo}}]{san03}
{S\'anchez Almeida}, J., {Emonet}, T., \& {Cattaneo}, F. 2003, \apj, 585, 536

\bibitem[{{S\'anchez Almeida} {et~al.}(1996){S\'anchez Almeida}, {Landi
  Degl'Innocenti}, {Mart\'\i nez Pillet}, \& {Lites}}]{san96}
{S\'anchez Almeida}, J., {Landi Degl'Innocenti}, E., {Mart\'\i nez Pillet}, V.,
  \& {Lites}, B.~W. 1996, \apj, 466, 537

\bibitem[{{S\'anchez Almeida} \& {Lites}(2000)}]{san00}
{S\'anchez Almeida}, J., \& {Lites}, B.~W. 2000, \apj, 532, 1215

\bibitem[{{S\'anchez Almeida} \& {Trujillo Bueno}(1999)}]{san99b}
{S\'anchez Almeida}, J., \& {Trujillo Bueno}, J. 1999, \apj, 526, 1013

\bibitem[{{Schrijver} \& {Title}(2003)}]{sch03b}
{Schrijver}, C.~J., \& {Title}, A.~M. 2003, \apjl, 597, L165

\bibitem[{{Sch\"ussler}(1986)}]{sch86}
{Sch\"ussler}, M. 1986, in Small Scale Magnetic Flux Concentrations in the
  Solar Photosphere, ed. W.~{Deinzer}, M.~{Kn\"olker}, \& H.~H. Voigt
  (G\"ottingen: Vandenhoeck \& Ruprecht), 103

\bibitem[{{Socas-Navarro} {et~al.}(2004){Socas-Navarro}, {Mart\'\i nez Pillet},
  \& {Lites}}]{soc04}
{Socas-Navarro}, H., {Mart\'\i nez Pillet}, V., \& {Lites}, B.~W. 2004, \apj,
  611, 1139

\bibitem[{{Socas-Navarro} \& {S\'anchez Almeida}(2002)}]{soc02}
{Socas-Navarro}, H., \& {S\'anchez Almeida}, J. 2002, \apj, 565, 1323

\bibitem[{{Socas-Navarro} \& {S\'anchez Almeida}(2003)}]{soc03}
---. 2003, \apj, 593, 581

\bibitem[{{Solanki}(1986)}]{sol86}
{Solanki}, S.~K. 1986, \aap, 168, 311

\bibitem[{{Solanki} \& {Steenbock}(1988)}]{sol88b}
{Solanki}, S.~K., \& {Steenbock}, W. 1988, \aap, 189, 243

\bibitem[{{Spruit}(1976)}]{spr76}
{Spruit}, H.~C. 1976, \solphys, 50, 269

\bibitem[{{Stein} \& {Nordlund}(2006)}]{stei06}
{Stein}, R.~F., \& {Nordlund}, {\AA}. 2006, \apj, 642, 1246

\bibitem[{{Stenflo}(1982)}]{ste82}
{Stenflo}, J.~O. 1982, \solphys, 80, 209

\bibitem[{{Trujillo Bueno} {et~al.}(2004){Trujillo Bueno}, {Shchukina}, \&
  Asensio~Ramos}]{tru04}
{Trujillo Bueno}, J., {Shchukina}, N.~G., \& Asensio~Ramos, A. 2004, \nat, 430,
  326

\bibitem[{{Unno}(1959)}]{unn59}
{Unno}, W. 1959, \apj, 129, 375

\bibitem[{{V{\" o}gler} {et~al.}(2005){V{\" o}gler}, {Shelyag}, {Sch{\"
  u}ssler}, {Cattaneo}, {Emonet}, \& {Linde}}]{vog05}
{V{\" o}gler}, A., {Shelyag}, S., {Sch{\" u}ssler}, M., {Cattaneo}, F.,
  {Emonet}, T., \& {Linde}, T. 2005, \aap, 429, 335

\bibitem[{{V\"ogler}(2003)}]{vog03b}
{V\"ogler}, A. 2003, PhD thesis, G\"ottingen University, G\"ottingen

\bibitem[{{V\"ogler} \& {Sch\"ussler}(2003)}]{vog03}
{V\"ogler}, A., \& {Sch\"ussler}, M. 2003, Astron. Nachr., 324, 399

\bibitem[{{V{\"o}gler} \& {Sch{\"u}ssler}(2007)}]{vog07}
{V{\"o}gler}, A., \& {Sch{\"u}ssler}, M. 2007, \aap, 465, L43

\bibitem[{{Yi} {et~al.}(1993){Yi}, {Jensen}, \& {Engvold}}]{yi93}
{Yi}, Z., {Jensen}, E., \& {Engvold}, O. 1993, in ASP Conf. Ser., Vol.~46, The
  Magnetic and Velocity Fields of Solar Active Regions, ed. H.~{Zirin},
  G.~{Ai}, \& H.~{Wang}, San Francisco, 232

\end{thebibliography}


\appendix
\section{Multi-component atmosphere with magnetic field vectors varying
in direction}
\label{appa}

The equations~(\ref{pdfIV}) and (\ref{pdfIVb}) used for multi-component
syntheses hold even when the magnetic field vector presents a
distribution of magnetic field inclinations, $\theta$, and  magnetic field 
azimuths, $\phi$. In the general case the PDF depends on $B$, $\theta$ and
$\phi$, $\mathcal{P}(B,\theta,\phi)$, so that the average Stokes $I$ and Stokes $V$
profiles are, 
\begin{equation}
I=\int_0^{2\pi}\int_0^\pi\int_{0}^{B_{\rm max}}\,\mathcal{P}(B,\theta,\phi)\,S_I(B,\theta,\phi)\, 
dB d\theta d\phi, 
\label{ap1}
\end{equation}
\begin{equation}
V=\int_0^{2\pi}\int_0^\pi\int_{0}^{B_{\rm max}}\,\mathcal{P}(B,\theta,\phi)\,S_V(B,\theta,\phi)\,
dB d\theta d\phi,
\label{ap2}
\end{equation}
with $\mathcal{P}(B,\theta,\phi)$ normalized to one.
The symbols $S_I(B,\theta,\phi)$ and $S_V(B,\theta,\phi)$ stand for the 
Stokes $I$ and $V$ profiles corresponding to each magnetic field vector.
As the tests carried out in \S~\ref{scs} show, and as it is to be expected
from the weakly polarizing medium approximation \citep{san99b}, 
the intensity of the weak Mn~{\sc i} lines is independent of inclination 
and azimuth so that $S_I(B,\theta,\phi)\simeq I_B$. Similarly, Stokes~$V$ is independent
of $\phi$ and depends on $\theta$ through the factor $\cos\theta$,
$S_V(B,\theta,\phi)=V_B\cos\theta$. The symbol $V_B$ represents the
Stokes~$V$ profile for longitudinal magnetic field 
($\cos\theta=1$).
Then from equations~(\ref{ap1}) and (\ref{ap2})
one retrieves the equations~(\ref{pdfIV}) and (\ref{pdfIVb}) used in the
main text,
\begin{equation}
I=\int_{0}^{B_{\rm max}}{P(B)\,I_B\,dB},
\end{equation}
\begin{equation}
V=\int_{0}^{B_{\rm max}}{\langle\cos\theta\rangle\,P(B)\,V_B\ dB},
\end{equation}
with
\begin{equation}
P(B)=\int_0^{2\pi}\int_0^\pi \mathcal{P}(B,\theta,\phi)\,d\theta d\phi, 
\end{equation}
and 
\begin{equation}
\langle\cos\theta\rangle ={1\over{P(B)}}
\int_0^{2\pi}\int_0^\pi\cos\theta\ \mathcal{P}(B,\theta,\phi)\,d\theta d\phi.
\end{equation}
\end{document}